\documentclass[a4paper,twocolumn]{aa}
\usepackage{amsmath,amssymb,amsfonts}
\usepackage[T1]{fontenc}
\usepackage{graphicx}
\usepackage[percent]{overpic}
\usepackage{tikz}
\usepackage{adjustbox,multirow}
\usepackage{hyperref}
\usepackage{natbib}
\bibpunct{(}{)}{;}{a}{}{,} 

\newcommand{\bb}[1]{{\boldsymbol{#1}}}

\renewcommand{\tt}[1]{{\boldsymbol{\widetilde #1}}}

\newcommand{\diag}{\mathrm{diag}}

\renewcommand{\d}{\mathrm{d}}

\newcommand{\e}{\mathrm{e}}
\newcommand{\G}{\mathcal{G}}

\renewcommand{\P}{\mathcal{P}}

\newcommand{\N}{\mathbb{N}}

\begin{document}

\title{The Denoised, Deconvolved, and Decomposed \emph{Fermi} $\gamma$-ray sky}
\subtitle{An application of the D$^\mathsf{3}$PO algorithm}

\titlerunning{The Denoised, Deconvolved, and Decomposed Fermi $\gamma$-ray sky}

\author{
    Marco~Selig\inst{\ref{inst1},\ref{inst2}}\and
    Valentina~Vacca\inst{\ref{inst1}}\and
    Niels~Oppermann\inst{\ref{inst3}}\and
    Torsten~A.~En{\ss}lin\inst{\ref{inst1},\ref{inst2}}
}

\institute{
    Max Planck Institut f\"ur Astrophysik (Karl-Schwarzschild-Stra{\ss}e~1, D-85748~Garching, Germany)\label{inst1}
    \and Ludwig-Maximilians-Universit\"at M\"unchen (Geschwister-Scholl-Platz~1, D-80539~M\"unchen, Germany) \label{inst2}
    \and Canadian Institute for Theoretical Astrophysics (60 St. George Street, Toronto, ON M5S 3H8, Canada) \label{inst3}
}

\date{Received DD MMM. YYYY / Accepted DD MMM. YYYY}

\abstract{
    We analyze the 6.5 year all-sky data from the \emph{Fermi} Large Area Telescope restricted to $\gamma$-ray photons with energies between 0.6--307.2 GeV.
    Raw count maps show a superposition of diffuse and point-like emission structures and are subject to shot noise and instrumental artifacts.
    Using the D$^3$PO inference algorithm, we model the observed photon counts as the sum of a diffuse and a point-like photon flux, convolved with the instrumental beam and subject to Poissonian shot noise. The D$^3$PO algorithm performs a Bayesian inference in this setting without the use of spatial or spectral templates; i.e., it removes the shot noise, deconvolves the instrumental response, and yields estimates for the two flux components separately.
    The non-parametric reconstruction uncovers the morphology of the diffuse photon flux up to several hundred GeV. 
    We present an all-sky spectral index map for the diffuse component.
    We show that the diffuse $\gamma$-ray flux can be described phenomenologically by only two distinct components: a soft component, presumably dominated by hadronic processes, tracing the dense, cold interstellar medium and a hard component, presumably dominated by leptonic interactions, following the hot and dilute medium and outflows such as the \emph{Fermi} bubbles.
    A comparison of the soft component with the Galactic dust emission indicates that the dust-to-soft-gamma ratio in the interstellar medium decreases with latitude. The spectrally hard component exists in a thick Galactic disk and tends to flow out of the Galaxy at some locations.
    Furthermore, we find the angular power spectrum of the diffuse flux to roughly follow a power law with an index of 2.47 on large scales, independent of energy.
    Our first catalog of source candidates includes $3{,}106$ candidates of which we associate $1{,}381$ ($1{,}897$) with known sources from the second (third) \emph{Fermi} source catalog. 
    We observe $\gamma$-ray emission in the direction of a few galaxy clusters hosting known radio halos. 
}

\keywords{methods: data analysis -- methods: statistical -- techniques: image processing -- gamma-rays: general -- gamma-rays: diffuse background -- catalogs}

\maketitle

\section{Introduction}

     \begin{figure*}[!t]
        \centering
        \begin{tabular}{cc}
            \begin{overpic} [width=0.47\textwidth]{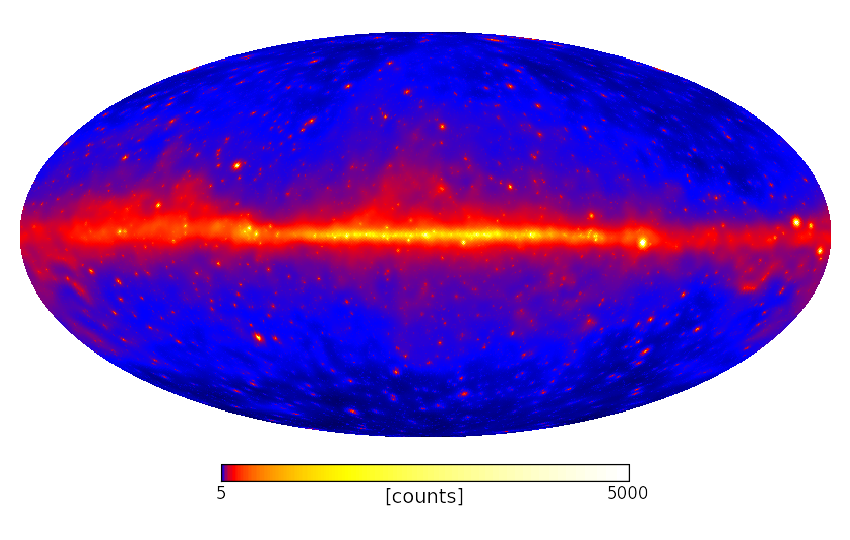} \put(-2,60){(a)} \end{overpic} &
            \begin{overpic} [width=0.47\textwidth]{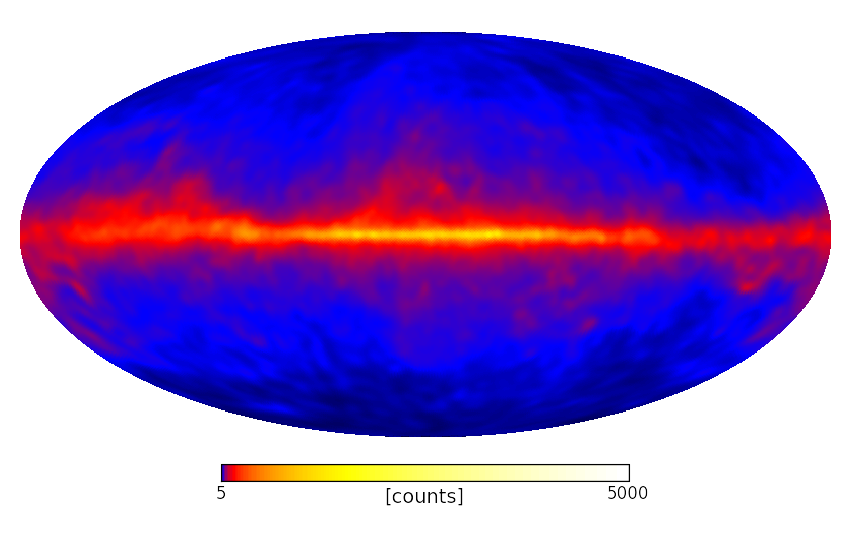} \put(-2,60){(b)} \end{overpic} \\
            \begin{overpic} [width=0.47\textwidth]{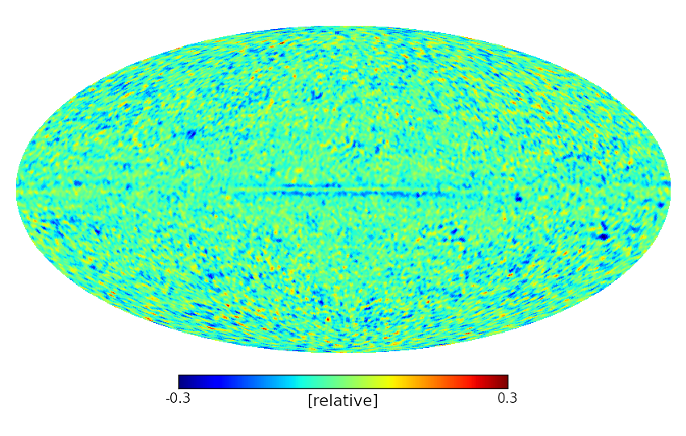} \put(-2,60){(c)$\,=\,$data$/$(a)$\,-\,1$} \end{overpic} &
            \begin{overpic} [width=0.47\textwidth]{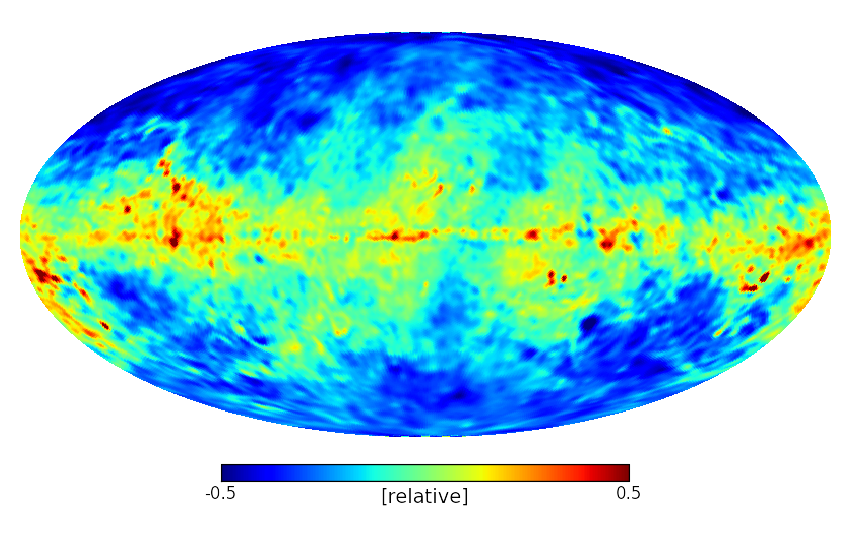} \put(-2,60){(d)$\,=\,$std$/$(b)$\,-\,1$} \end{overpic} \\
        \end{tabular}
        \flushleft
        \caption{Illustration of the $\gamma$-ray sky seen by the \emph{Fermi} LAT in Mollweide projection. Panel (a) shows the (total) photon flux reconstructed from photon count data of $6.5$ years mission elapsed time in the energy range from $0.6$ to $307.2 \, \mathrm{GeV}$ reconvolved with the LAT's IRFs. Panel (b) shows solely the reconvolved diffuse contribution. Panel (c) shows the fractional residual map between data and reconstruction smoothed with a $0.5^\circ$ Gaussian kernel. Panel (d) shows the fractional residual map between the standard Galactic diffuse model (short ``std'') and the reconstructed diffuse contribution smoothed with a $0.75^\circ$ Gaussian kernel.}
        \label{fig:sum}
    \end{figure*}

    Since August 2008 the \emph{Fermi} Gamma-ray Space Telescope has observed the $\gamma$-ray sky with its main instrument, the Large Area Telescope \citep[LAT,][]{F09a}, which is sensitive to photons with energies ranging from around $20 \, \mathrm{MeV}$ to above $300 \, \mathrm{GeV}$.

    There is a diversity of astrophysical contributions to the total $\gamma$-ray flux.
    Most of the photons in the $\mathrm{GeV}$-range are induced by cosmic rays (CRs), charged particles moving at (ultra-)relativistic speeds, through hadronic interactions of CR nuclei with the interstellar medium (ISM) or inverse Compton scattering (IC) of electrons with background light \citep{F12a,D13}. In addition, there is emission from an spatially constant diffuse background, which is commonly denoted as ``extragalactic'' background \citep[][and references therein]{D07}, and from sources that appear point-like.

    The diffuse and point-like $\gamma$-ray fluxes appear superimposed to an observer. An observation through an instrument, like the \emph{Fermi} LAT for example, additionally convolves the total flux with the instrument response functions (IRFs). The gathered data are, lastly, subject to noise; i.e., Poissonian shot noise in the case of integer photon counts.
    In order to retrieve the physical photon flux from observations, we would need to reverse those processes.
    Unfortunately, neither a direct inversion of the convolution nor the separation of noise and signal components is feasible exactly, so that we have to resort to alternative approaches.

    One possibility is ``forward'' modeling, whereby parametrized models of different emission components are fit to the data; e.g., by a maximum-likelihood procedure as suggested by \citet{F08} for the analysis of \emph{Fermi} LAT data.
    By inspection of residuals between the data and the best fitting model(s), new features might be discovered.
    A famous example are the \emph{Giant Fermi Bubbles} revealed by \citet{SSF10} using templates. Such templates are commonly constructed in accordance with surveys at lower energies or by modeling of the relevant CR physics \citep[][and references therein]{F08,SSF10,SF12,F12a,F14a}.

    In this work, we investigate the ``backward'' reconstruction of flux contributions using Bayesian inference methods \citep{B63,C46,S48,W49}. The idea is to obtain signal estimates (and uncertainties) from an algorithm based on a probabilistic model that denoises, deconvolves, and decomposes the input data. This probabilistic model includes prior constraints to remedy the complexity of the inverse problem.
    Assuming a sparsity-based regularization, for example, \citet{S10,S12} proposed an analysis strategy using waveforms, which they applied to simulated \emph{Fermi} data.
    For the analysis of X-ray images, which pose the same challenges as $\gamma$-ray images, a Bayesian background-source separation technique was proposed by \citet{GFD09}.

    We deploy the D$^3$PO inference algorithm \citep{SE13} derived within the framework of information field theory \citep[IFT,][]{EFK09,E13,E14}. It simultaneously provides non-parametric estimates for the diffuse and the point-like photon flux given a photon count map.
    This challenging inverse problem is thereby regularized by prior assumptions that provide a statistical description of the morphologically different components; i.e., the priors define our naive understanding of ``diffuse'' and ``point-like''.
    D$^3$PO considers Poissonian shot noise, without Gaussian approximations, and takes the provided IRFs of the \emph{Fermi} LAT fully into account. Furthermore, we can retrieve uncertainty information on the estimates.

    All this allows us to present a continuous reconstruction of the diffuse $\gamma$-ray sky up to around $300 \, \mathrm{GeV}$, as well as an estimate of the point-like contribution, from which we derive the first D$^3$PO \emph{Fermi} catalog of $\gamma$-ray source candidates.
    By analyzing the spectral behavior of the diffuse component, it is possible to investigate the underlying processes, especially with regard to the CRs responsible for the emission.

    The remainder of this paper is structured as follows. Section~\ref{sec:anasum} summarizes the analysis procedure, a more detailed description is given in Appendix~\ref{app:ana}. We present and discuss our findings in Section~\ref{sec:discussion}, and conclude in Section~\ref{sec:conclusion}.

\section{Analysis summary}
\label{sec:anasum}

    We analyze the photon count data collected by the \emph{Fermi} LAT within its $6.5$ years of operation. All data selection criteria, the working principle of the applied D$^3$PO inference algorithm \citep{SE13}, and a description of the analysis procedure are detailed in Appendix~\ref{app:ana}.

    In summary, we make use of the reprocessed Pass 7 available within the \emph{Fermi} Science Tools\footnote{\url{http://fermi.gsfc.nasa.gov/ssc/data/analysis/documentation/}} in order to retrieve the data, as well as the corresponding instrument response functions and exposure of the \emph{Fermi} LAT \citep{F09a,F09c,F12c}.
    We consider nine logarithmically spaced energy bands ranging from $0.6$ to $307.2 \, \mathrm{GeV}$, cf. Table~\ref{tab:E}. For each band, we spatially bin all events classified as \texttt{CLEAN} in count maps, whereby we distinguish the front or back conversion of the photon within the LAT. Throughout this work, we discretize the sky using the \textsc{HEALPix} scheme with $n_\mathrm{side} = 128$, which corresponds to an angular resolution of approximately $0.46^\circ$.

    \begin{figure*}[!t]
        \centering
        \begin{tabular}{ccc}
            \begin{overpic} [width=0.3\textwidth]{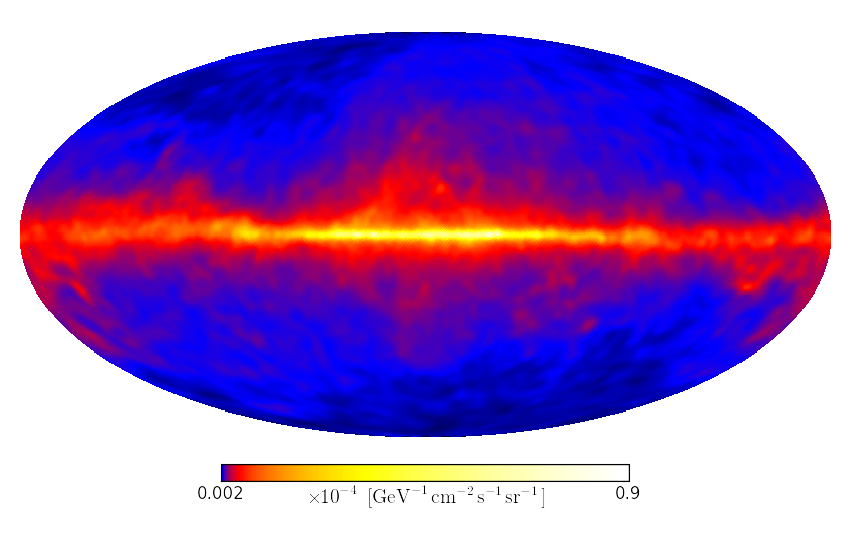} \put(-5,61){(a)} \end{overpic} &
            \begin{overpic} [width=0.3\textwidth]{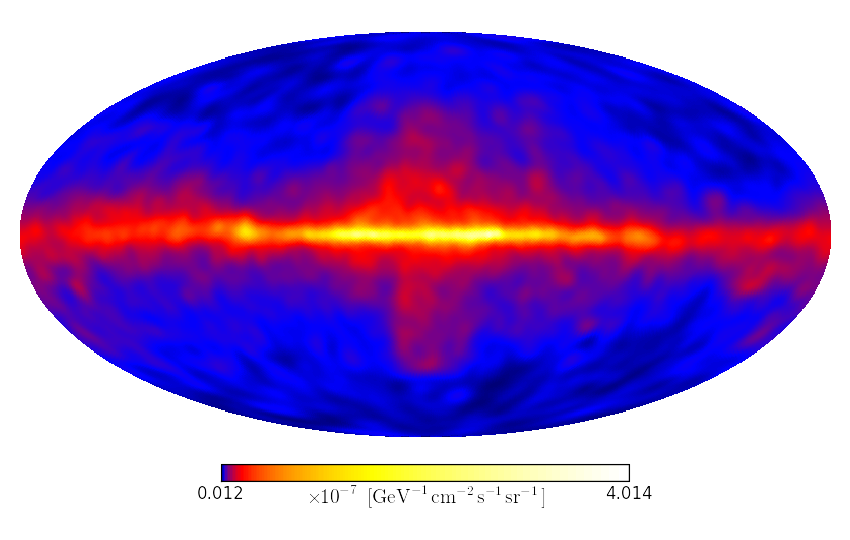} \put(-5,61){(b)} \end{overpic} &
            \begin{overpic} [width=0.3\textwidth]{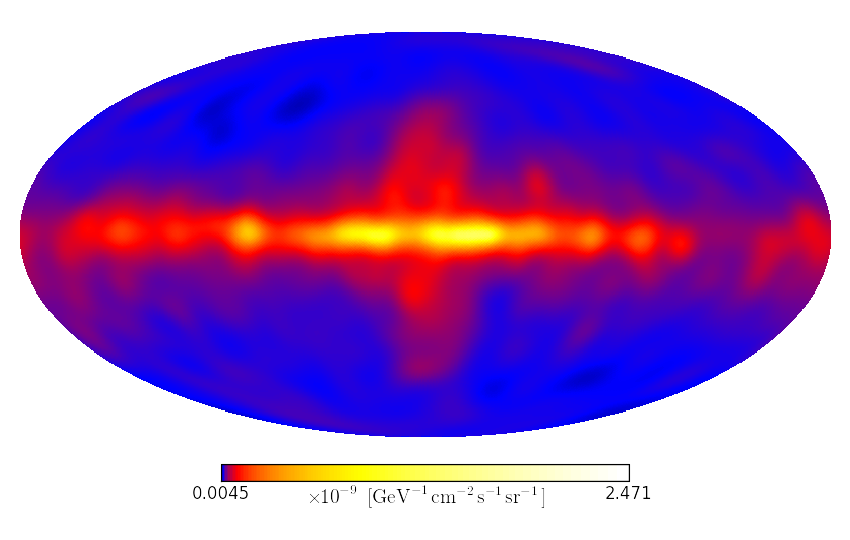} \put(-5,61){(c)} \end{overpic} \\
            \begin{overpic} [width=0.3\textwidth]{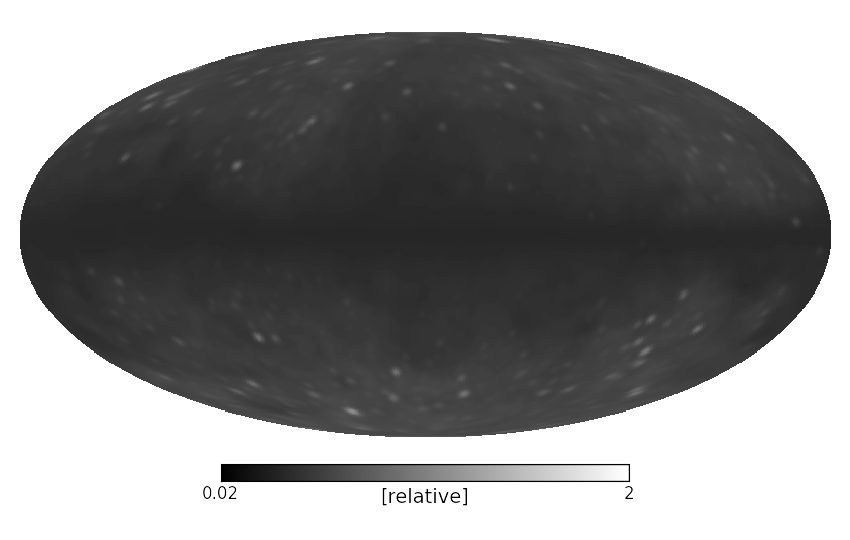} \put(-5,61){(d)} \end{overpic} &
            \begin{overpic} [width=0.3\textwidth]{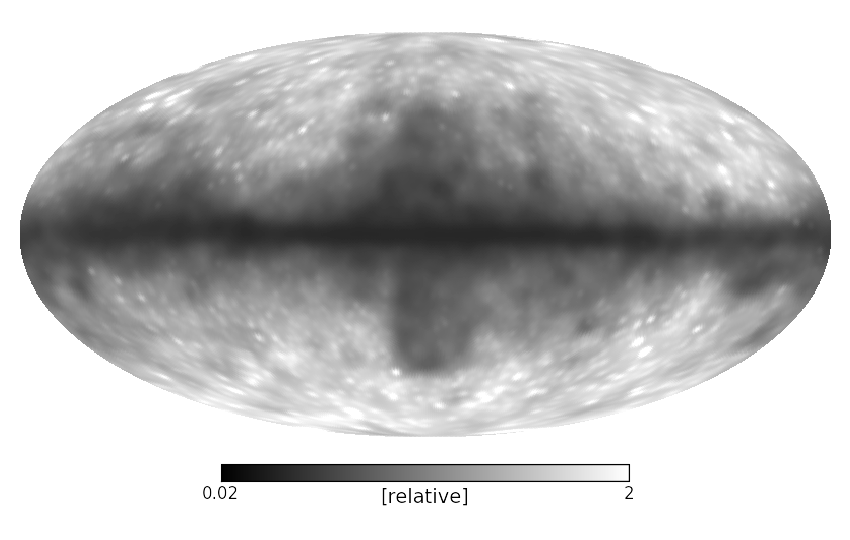} \put(-5,61){(e)} \end{overpic} &
            \begin{overpic} [width=0.3\textwidth]{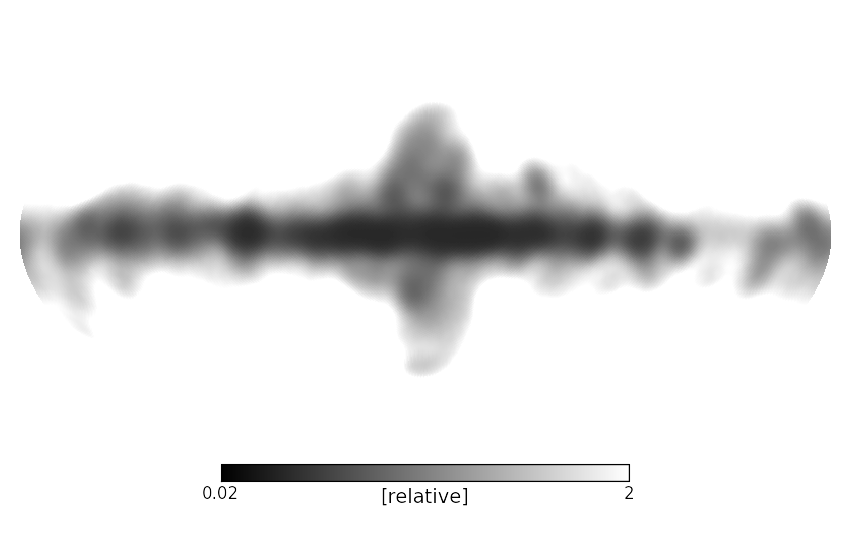} \put(-5,61){(f)} \end{overpic} \\
        \end{tabular}
        \flushleft
        \caption{Illustration of the diffuse $\gamma$-ray flux and its relative reconstruction uncertainty on a logarithmic scale at different energies in Mollweide projection. The panels (a--c) show the reconstructed diffuse photon flux at roughly $2 \, \mathrm{GeV}$, $10 \, \mathrm{GeV}$, and $\sim 100 \, \mathrm{GeV}$. The photon flux is given in units of $\mathrm{GeV}^{-1} \mathrm{cm}^{-2} \mathrm{s}^{-1} \mathrm{sr}^{-1}$. Notice that the color scale varies. The panels (d--f) show the relative uncertainty on the above reconstructions. Maps for all energy bands are contained in the online material.}
        \label{fig:mapsplus}
    \end{figure*}

    \begin{table}[!b]
        \caption{Overview of the energy binning. Listed are minimum, logarithmic mean, and maximum energy for each bin.}
        \centering
        \begin{tabular}{|crrr|}
            \hline
            band $^{\phantom{'}}$ & \multicolumn{1}{c}{$E^\mathrm{min} \,[\mathrm{GeV}]$} & \multicolumn{1}{c}{$E^\mathrm{mid} \,[\mathrm{GeV}]$} & \multicolumn{1}{c|}{$E^\mathrm{max} \,[\mathrm{GeV}]$} \\
            \hline
            \hline
            $^{\phantom{'}}1^{\phantom{'}}$ &   $0.60$ &   $0.85$ &   $1.20$ \\
            $2$ &   $1.20$ &   $1.70$ &   $2.40$ \\
            $3$ &   $2.40$ &   $3.40$ &   $4.80$ \\
            $4$ &   $4.80$ &   $6.79$ &   $9.60$ \\
            $5$ &   $9.60$ &  $13.58$ &  $19.20$ \\
            $6$ &  $19.20$ &  $27.15$ &  $38.40$ \\
            $7$ &  $38.40$ &  $54.31$ &  $76.80$ \\
            $8$ &  $76.80$ & $108.61$ & $153.60$ \\
            $9$ & $153.60$ & $217.22$ & $307.20$ \\
            \hline
        \end{tabular}
        \label{tab:E}
    \end{table}

    This data set is the input for the D$^3$PO algorithm. In order to denoise, deconvolve, and decompose the data, we suppose the data $\bb{d}$ to be the result of a Poisson process with an expectation value given by the convolved sum of the diffuse and point-like flux, $\bb{\phi}^{(s)}$ and $\bb{\phi}^{(u)}$; i.e.,
    \begin{align}
        \bb{d} \curvearrowleft \P(\bb{d}|\bb{R}(\bb{\phi}^{(s)}+\bb{\phi}^{(u)}))
        ,
    \end{align}
    where the operator $\bb{R}$ describes the full instrument response of the LAT.
    Under this model assumption, and with the aid of prior regularizations, D$^3$PO computes estimates for the photon fluxes.
    The inference is performed iteratively until convergence and for each energy band separately.
    Further details regarding the inference can be found in Appendix~\ref{app:ana}.

    The results of this analysis -- including, among others, the reconstructed fluxes, uncertainties, and the first D$^3$PO \emph{Fermi} (1DF) catalog of $\gamma$-ray source candidates -- are publicly available at \url{http://www.mpa-garching.mpg.de/ift/fermi/}.

\section{Results and discussion}
\label{sec:discussion}

    \begin{figure*}[!t]
        \centering
        \begin{tabular}{ccc}
            \begin{overpic} [width=0.3\textwidth]{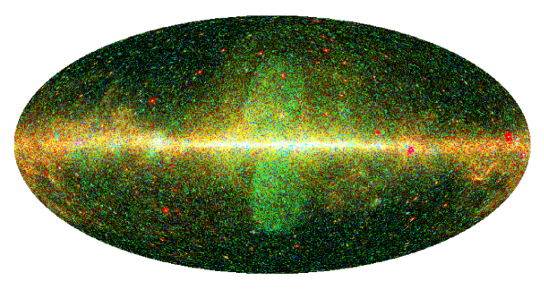} \put(-4,51){(a)} \end{overpic} &
            \begin{overpic} [width=0.3\textwidth]{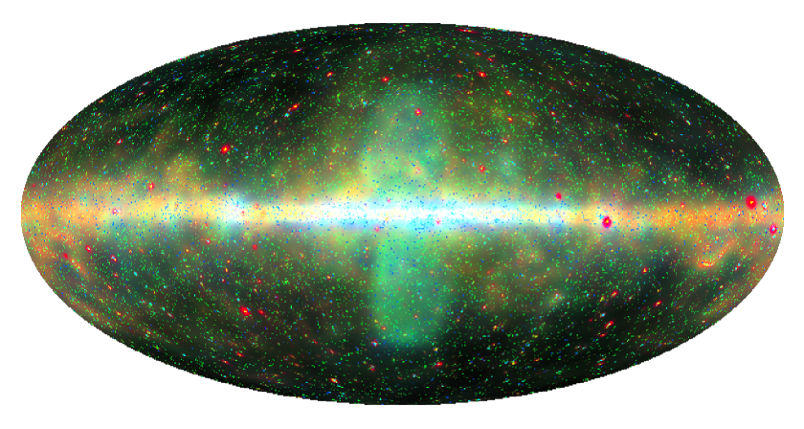} \put(-4,51){(b)} \end{overpic} &
            \begin{overpic} [width=0.3\textwidth]{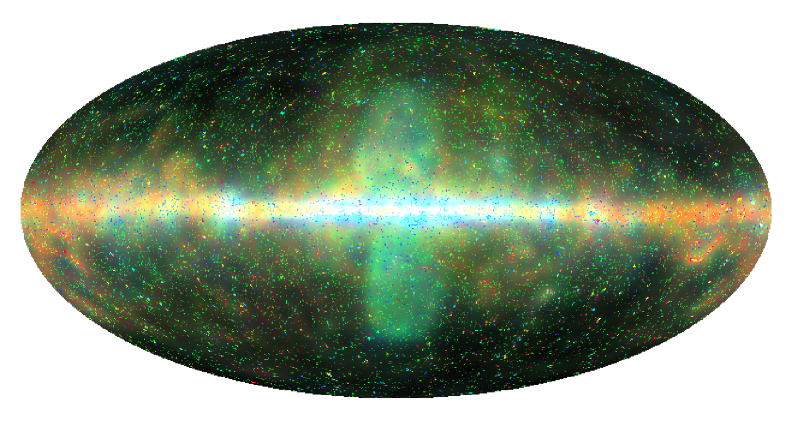} \put(-4,51){(c)} \end{overpic} \\
        \end{tabular}
        \begin{overpic} [width=\textwidth]{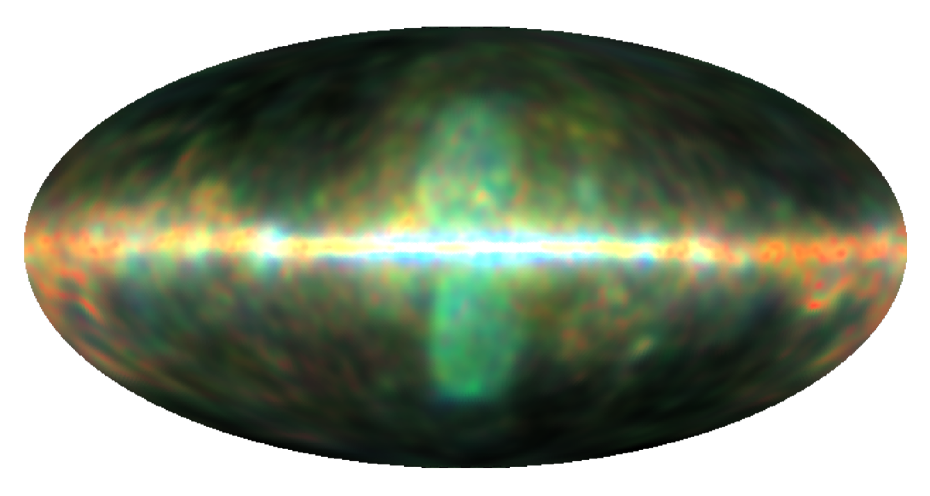} \put(1.4,51){(d)} \end{overpic}
        \begin{tabular}{ccc}
            \begin{overpic} [width=0.3\textwidth]{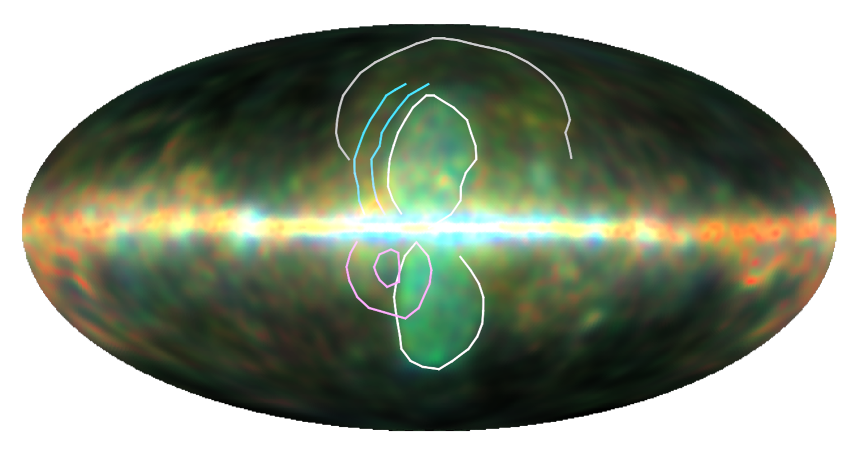} \put(-4,51){(e)} \end{overpic} &
            \begin{overpic} [width=0.3\textwidth]{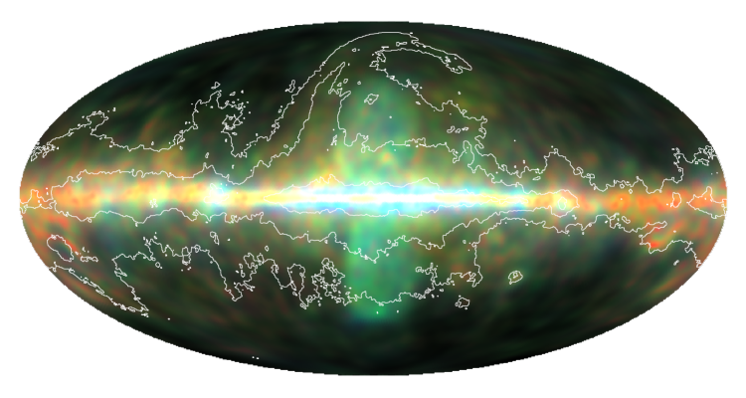} \put(-4,51){(f)} \end{overpic} &
            \begin{overpic} [width=0.3\textwidth]{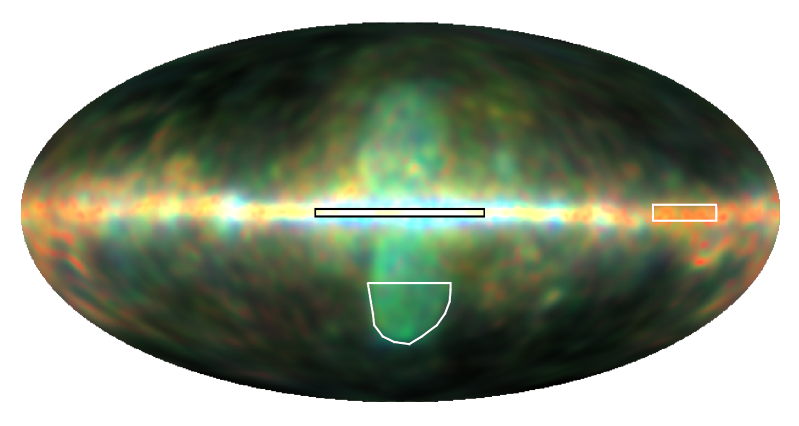}     \put(-4,51){(g)} \end{overpic} \\
        \end{tabular}
        \flushleft
        \caption{Illustration of the $\gamma$-ray sky in pseudocolor in Mollweide projection. Panel (a) shows the $6.5$ year data. The panels (b) and (c) show the reconstructed (total) photon flux that in (b) is reconvolved with the IRFs. Panel (d) shows the reconstructed diffuse photon flux. The panels (e) and (f) reproduce the latter but are overlaid with the feature contours found by \citet{SSF10} (white: \emph{Giant Fermi Bubbles}, light magenta: \emph{Donut}, light blue: \emph{North Arc}, light gray: \emph{Radio Loop I}) and contours of the $408 \, \mathrm{MHz}$ radio map from \citet{HSS82}, respectively.
        Panel (g) highlights the contours defining the ``bulge''-, ``cloud''-, and ``bubble''-like regions discussed in Section~\ref{sec:like}.}
        \label{fig:pseudo}
    \end{figure*}

\subsection[The gamma-ray Sky]{The $\gamma$-ray sky}

    The \emph{Fermi} LAT has detected millions of $\gamma$-ray photons within the first $6.5$ years of its mission.
    We can stack them in a binned all-sky count map disregarding the energy of the photons.

    Figure~\ref{fig:sum} shows such all-sky maps for the total and the diffuse result of the reconstruction. D$^3$PO denoises, deconvolves, and decomposes photon count maps yielding a reconstruction of the diffuse and point-like photon flux. A reconvolution of the reconstructed $\gamma$-ray flux with the instrument response functions demonstrates the quality of the performed denoising; i.e., the removal of Poissonian shot noise.

    The fractional residual exhibits no significant structures and the mean of its absolute value is below $5\%$.\footnote{If we approximate the Poissonian $\P(\bb{d}|\bb{\lambda})$ by a Gaussian $\G$ with mean $\bb{\lambda}$ and variance $\diag[\bb{\lambda}]$, we can expect the mean of the absolute fractional residual to be
    \begin{align}
        \frac{1}{N} \sum_{n=1}^N \frac{|d_n-\lambda_n|}{\lambda_n} \approx \frac{1}{N} \sum_{n=1}^N \frac{|\G(0,1)|}{\sqrt{\lambda_n}} \approx \frac{1}{N} \sum_{n=1}^N \sqrt{\frac{2}{\pi d_n}}. \nonumber
    \end{align}
    For the considered data set, this computes to around $16\%$.}
    As we compare the data with a denoised reproduction, the major difference is due to shot noise.
    We observe a weak ringing around the Galactic plane, which is a numerical artifact due to imperfections of spherical harmonic transformations applied during the inference.
    Previous comparisons with best-fitting templates created by the GALPROP code \citep[][and references therein]{MS00,SMR00,F08,F12a} often show significant residuals indicating features lacked by the respective models; e.g., cf. Fig. 6 in \citet{F12a}. Since our inference machinery, on the contrary, is free of \emph{a~priori} assumption regarding the existence of any Galactic or extragalactic features, significant residuals are not to be expected.

    Excluding the point-like contribution from the reconvolved count map, the diffuse $\gamma$-ray sky becomes fully revealed, see Fig.~\ref{fig:sum}b. The diffuse count map clearly displays Galactic features and substructures within the ISM.
    In comparison to the standard Galactic diffuse model\footnote{The standard Galactic diffuse model is provided by \url{http://fermi.gsfc.nasa.gov/ssc/data/analysis/software/aux/gll_iem_v05_rev1.fit}.}, we find obvious residuals. These include diffuse structures on very small scales that are not captured in the reconstruction because its effective resolution is limited due to signal-to-noise and IRFs. At high latitudes, extragalactic contributions are reconstructed and yield an excess compared to the Galactic diffuse model.
    While the reconvolved photon count image appears somewhat smoothed, its deconvolved counterpart displays the Milky Way in more detail. The diffuse $\gamma$-ray fluxes in the individual energy bands are shown in Fig.~\ref{fig:mapsplus}. The coarseness of the images increases with energy because the number of detected photons, and thus the signal-to-noise ratio, drops drastically. The uncertainties of the reconstructions are illustrated in the lower panels of Fig.~\ref{fig:mapsplus}.
    Nevertheless, the Galactic disk and bulge are clearly visible at all energies.

\subsubsection{Pseudocolor images}
\label{sec:pseudocolor}

    In order to obtain a better view on the spectral characteristics of the $\gamma$-ray sky, we combine the maps at different energies by a pseudocolor scheme. This scheme is designed to mimic the human perception of optical light in the $\gamma$-ray range. Intensity indicates the (logarithmic) brightness of the flux, and red colors correspond to low energy $\gamma$-rays around $1 \, \mathrm{GeV}$ and blue colors to $\gamma$-rays up to $300 \, \mathrm{GeV}$.
    The resulting pseudocolor maps of the $\gamma$-ray sky are presented in Fig.~\ref{fig:pseudo}.
    Thanks to a suitably tuned color response, spectrally different regions can easily be identified by the human eye. At a first glance, we can recognize the bright bulge of the Milky Way, the \emph{Fermi} bubbles as two greenish blue, roundish areas, and red to yellowish cloud-like structures at low and intermediate latitudes, in particular around the Galactic anticenter.

    The upper panels (a--d) illustrate the functionality of the D$^3$PO inference algorithm showing the raw data and the denoised, deconvolved, and decomposed reconstruction, respectively. The denoising applies most strongly to the high energy bands, appearing green- to blueish, where the signal-to-noise ratios are worst. The deconvolution effect is most evident for point-like contributions in lower energy bands, appearing reddish, because of the increasing width of the point spread function (PSF) for these bands. Finally, the decomposition reveals the purely diffuse $\gamma$-ray sky.

    This view reveals many interesting features beyond the Galactic disk and bulge, which we will discuss in the following.

\subsubsection{Bubbles, features, and radio}

    The most striking features recovered by our reconstruction are the \emph{Giant Fermi Bubbles} first found by \citet{SSF10}. The bubbles extend up to $|b| \lesssim 50^\circ$ in latitude and $|l| \lesssim 20^\circ$ in longitude. They appear to emerge from the Galactic center, however, their astrophysical origin is still under discussion \citep[][and references therein]{SSF10,CA11,C+11,D+11,SF12,YAC14,F14a}.
    In agreement with previous studies, we find the bubbles to have relatively sharp edges and an overall homogeneous surface brightness, appearing greenish blue in Fig.~\ref{fig:pseudo}. \citet{YAC14} report an energy dependent morphology of the southern bubble, which is, in particular, more extended to the Galactic South and West at high energies. Our results confirm this extension, as can be seen in the reconstruction for the highest energy band in Fig.~\ref{fig:mapsplus}.

    Figure~\ref{fig:pseudo}e also shows the \emph{North Arc}, \emph{Donut} and \emph{Cocoon} \citep{SSF10,SF12}. However, we do not find evidence for a jet-like structure as reported by \citet{SF12}.


    Moreover, there is a correlation with structures seen at radio frequencies. For example, a comparison with the synchrotron map from \citet{HSS82} taken at $408 \, \mathrm{MHz}$ reveals $\gamma$-ray counterparts of the \emph{Radio Loop I} \citep{LQH62} and smaller objects like the \emph{Large Magellanic Cloud} at $(l,b) \approx (-80^\circ,-30^\circ)$, as well as the $\gamma$-ray glow around \emph{Centaurus A} at $(l,b) \approx (-50^\circ,20^\circ)$, see Fig.~\ref{fig:pseudo}f.
    However, the resolution of the all-sky reconstruction is too coarse to detail the morphology of such small sources. A reconstruction of a more focused field of view would be necessary to that end.

    \begin{figure}[!t]
        \centering
        \begin{overpic} [width=0.5\textwidth]{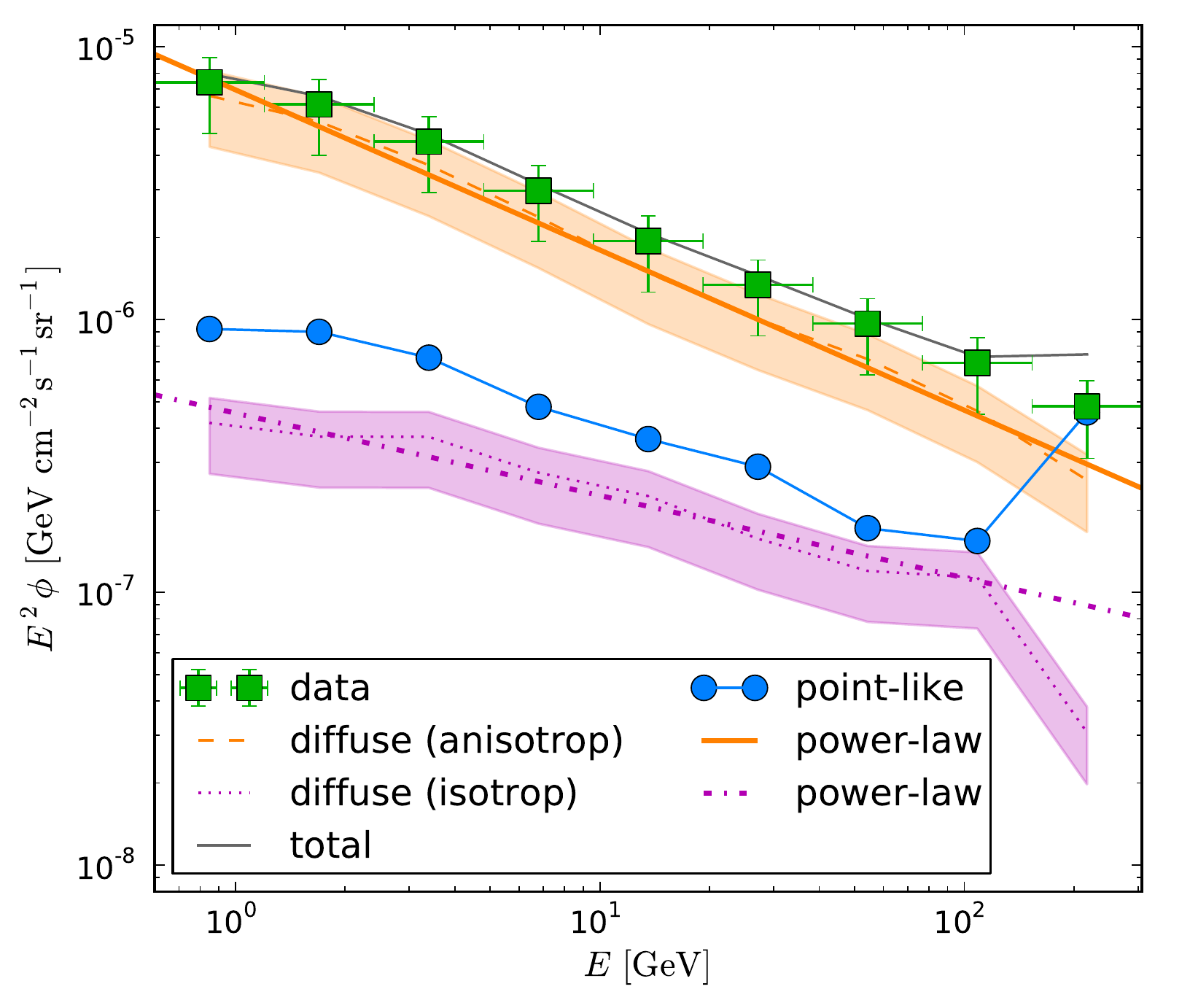} \put(75,77){{\tiny ROI: all-sky}} \end{overpic}
        \flushleft
        \caption{Illustration of energy spectra considering an all-sky ROI. Shown are the data (green squares) converted to flux units, and spectra from the reconstructed total (gray), anisotropic diffuse (dashed orange), isotropic diffuse (dotted magenta), and point-like photon flux (blue circles). Furthermore, power-law fits for the anisotropic (thick solid orange) and isotropic emission (thick dash-dotted magenta) are shown. The errors include statistical and systematic uncertainties and are only shown for data and diffuse contributions for reasons of clarity.}
        \label{fig:speck}
    \end{figure}

    \begin{figure}[!b]
        \centering
        \includegraphics[width=0.47\textwidth]{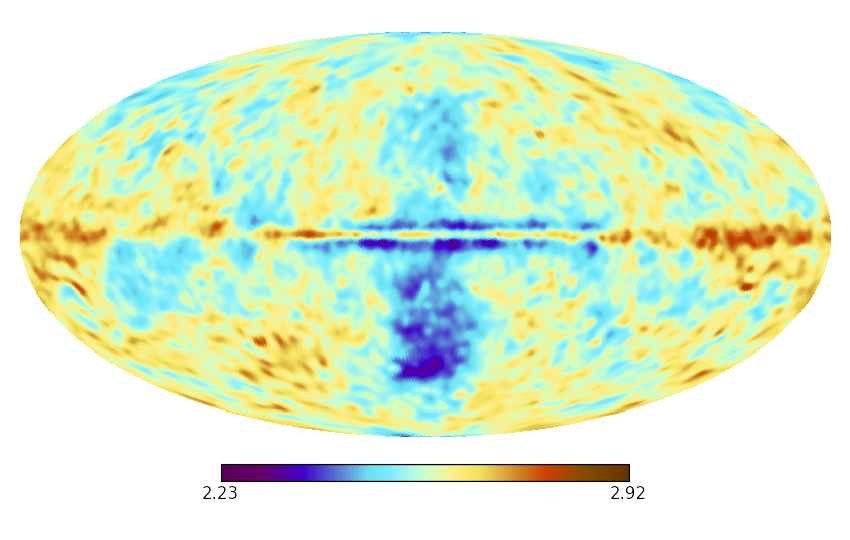}
        \flushleft
        \caption{Illustration of the spectral indices of featureless power-law fits, $\phi_\mathrm{aniso} \propto E^{-\gamma}$, at all positions in the sky using energy bands $1$--$8$.}
        \label{fig:speckmap}
    \end{figure}

\subsection{Energy spectra and spectral indices}
\label{sec:speck}

    In order to get a more quantitative view on the different contributions to the $\gamma$-ray flux, we now investigate photon flux energy spectra.

    Figure~\ref{fig:speck} shows the measured\footnote{The photon data $\bb{d}$ can be converted into flux units by division by the respective exposure $\varepsilon$, solid angle $\Omega$, and width of the energy band, according to
    \begin{align}
        \phi_j = \phi(E_j^\mathrm{mid}) &\equiv \frac{1}{(E_j^\mathrm{max}-E_j^\mathrm{min})} \sum_{i \in \mathrm{ROI}} \frac{1}{2\Omega^\mathrm{ROI}} \left( \frac{d_{ij}^{\,\texttt{FRONT}}}{\varepsilon_{ij}^{\,\texttt{FRONT}}} + \frac{d_{ij}^{\,\texttt{BACK}}}{\varepsilon_{ij}^{\,\texttt{BACK}}} \right)
        , \nonumber 
    \end{align}
    where the indices $i$ and $j$ label pixels and energy bands, respectively. Notice that front- and back- converted data are averaged accordingly.} and reconstructed energy spectra for the whole sky. Further regions of interest (ROIs), which typically in- or exclude the Galactic center or bulge, are investigated as well, see results in Table~\ref{tab:roi}.
    The errors are dominated by systematics; i.e., by the uncertainty in the absolute energy scale $\Delta \mathcal{E}/\mathcal{E} = (+2\%,-5\%)$ \citep{F12e} and in the normalization of the effective area, which is $\pm 10\%$ up to $100 \, \mathrm{GeV}$ and increases linearly with $\log(E)$ to $\pm 15\%$ at $1 \, \mathrm{TeV}$ \citep{F13b}\footnote{See also \url{http://fermi.gsfc.nasa.gov/ssc/data/analysis/LAT_caveats.html}.}. The statistical uncertainties determined from the inference tend to track the signal-to-noise ratio.

    We split the total energy spectrum into a diffuse and a point-like contribution, whereby we additionally distinguish between isotropic\footnote{In this context, ``isotropic'' means spatially constant.} and anisotropic diffuse components. Overall, the spectra from the reconstructed fluxes agree well with the data, except for the highest energy bin, where the point-like component seems to be strongly overestimated.
    There are two reasons for this. On the one hand, the signal-to-noise ratio is lowest, and on the other hand, the PSF is sharpest. Therefore, the distinction between point-like sources, noise peaks, and weak diffuse emission breaks down. For this reason, we exclude this highest energy band from further spectral analysis.

    The diffuse $\gamma$-ray flux amounts to $\sim 90\%$ of the total flux with the majority being anisotropic contributions of Galactic origin.
    Both diffuse contributions, isotropic and anisotropic, are consistent with featureless power laws, $\phi \propto E^{-\gamma}$.
    The results of the power-law fits are given in Table~\ref{tab:roi}.

    For the anisotropic component we find a spectral index $\gamma^{(s)} = 2.59 \pm 0.05$ that is slightly softer than the index of $2.44 \pm 0.01$ reported by \citet{F12b} indicating a lack of flux at high energies, which can again be explained by the low signal-to-noise ratio in this regime.
    In low energy bands, the spectrum is dominated by the production and decay of $\pi^0$-mesons induced by CR protons, while IC emission becomes increasingly important at the highest energies \citep{F12b,F12a}.
    The declining tail of the pion bump, peaking at $\tfrac{1}{2} m_{\pi^0} \approx 0.07 \, \mathrm{GeV} \, \mathrm{c}^{-2}$, is visible.

    The isotropic background is often referred to as ``extragalactic'' because it comprises unresolved extragalactic sources and might include possible signatures from large-scale structure formation or dark matter decay \citep[][and references therein]{D07,F10a}.
    The isotropic diffuse background also follows a featureless power law with a spectral index $\gamma_\mathrm{iso}^{(s)} = 2.30 \pm 0.06$, if we ignore the last energy band.
    Notice that the excess of isotropic emission around $\sim 3 \, \mathrm{GeV}$ is rather insignificant with regard to the uncertainties.\footnote{Such a tentative excess could be a line-like signal from dark matter decay/annihilation \citep{C12}, however, this is highly speculative considering the available data.}
    \citet{F10a} derive a spectral index of $2.41 \pm 0.05$ from $1$ year \emph{Fermi} LAT data in the energy range of $0.03$--$100 \, \mathrm{GeV}$. This indicates a slight spectral hardening of the isotropic background towards higher energies.
    In the same energy range, observations with the Energetic Gamma Ray Experiment Telescope (EGRET) yield a spectral index of $2.10 \pm 0.03$ that is considerably smaller \citep{S98}. This discrepancy, which might be an instrumental issue, is not yet clarified.

    The recent analysis by \citet{F14b} investigating an energy range of $0.1$--$820 \, \mathrm{GeV}$ reports the isotropic $\gamma$-ray background to be consistent with a power law with exponential cut-off at $280 \, \mathrm{GeV}$ having a spectral index of $2.32 \pm 0.02$. This is consistent with our findings.
\begin{table*}[!t]
        \caption{Overview of the ROIs. Listed are inclusion cuts in Galactic longitude $l$ and latitude $b$, covered solid angle $\Omega$, fitted spectral indices $\gamma^{(s)}$, and $\chi^2$ divided by the degrees of freedom (DOF); i.e., $9$ bands$-2$ unknowns $=7$. The latter two are given for the anisotropic and the isotropic diffuse photon flux, respectively.}
        \centering
        \begin{tabular}{|clccccc|}
            \hline
            ROI $^{\phantom{'}}$ & \multicolumn{1}{c}{inclusion cuts} & $\Omega \,[\mathrm{sr}]$ & $\gamma^{(s)}$ & $\chi^2/\mathrm{DOF}$ & $\gamma_\mathrm{iso}^{(s)}$ & $\chi_\mathrm{iso}^2/\mathrm{DOF}$ \\
            \hline
            \hline
            $^{\phantom{'}}1^{\phantom{'}}$ & \multicolumn{1}{c}{all-sky} & $\phantom{0}4\pi$ & $2.59 \pm 0.05$ & $0.10$ & $2.30 \pm 0.06$ & $0.14$ \\
            $2$ & $|b|>10^\circ \;\lor\; |l|<10^\circ$ & $10.5$                               & $2.60 \pm 0.05$ & $0.22$ & $2.30 \pm 0.06$ & $0.14$ \\
            $3$ & $|b|>10^\circ$ & $10.4$                                                     & $2.63 \pm 0.05$ & $0.39$ & $2.30 \pm 0.06$ & $0.14$ \\
            $4$ & $|b|<10^\circ \;\land\; |l|<80^\circ$ & $\phantom{0}4.5$                    & $2.52 \pm 0.05$ & $0.08$ & $2.52 \pm 0.06$ & $0.06$ \\
            \hline
        \end{tabular}
        \label{tab:roi}
    \end{table*}

    \begin{figure*}[!t]
        \centering
        \begin{tabular}{cc}
            \begin{overpic} [width=0.5\textwidth]{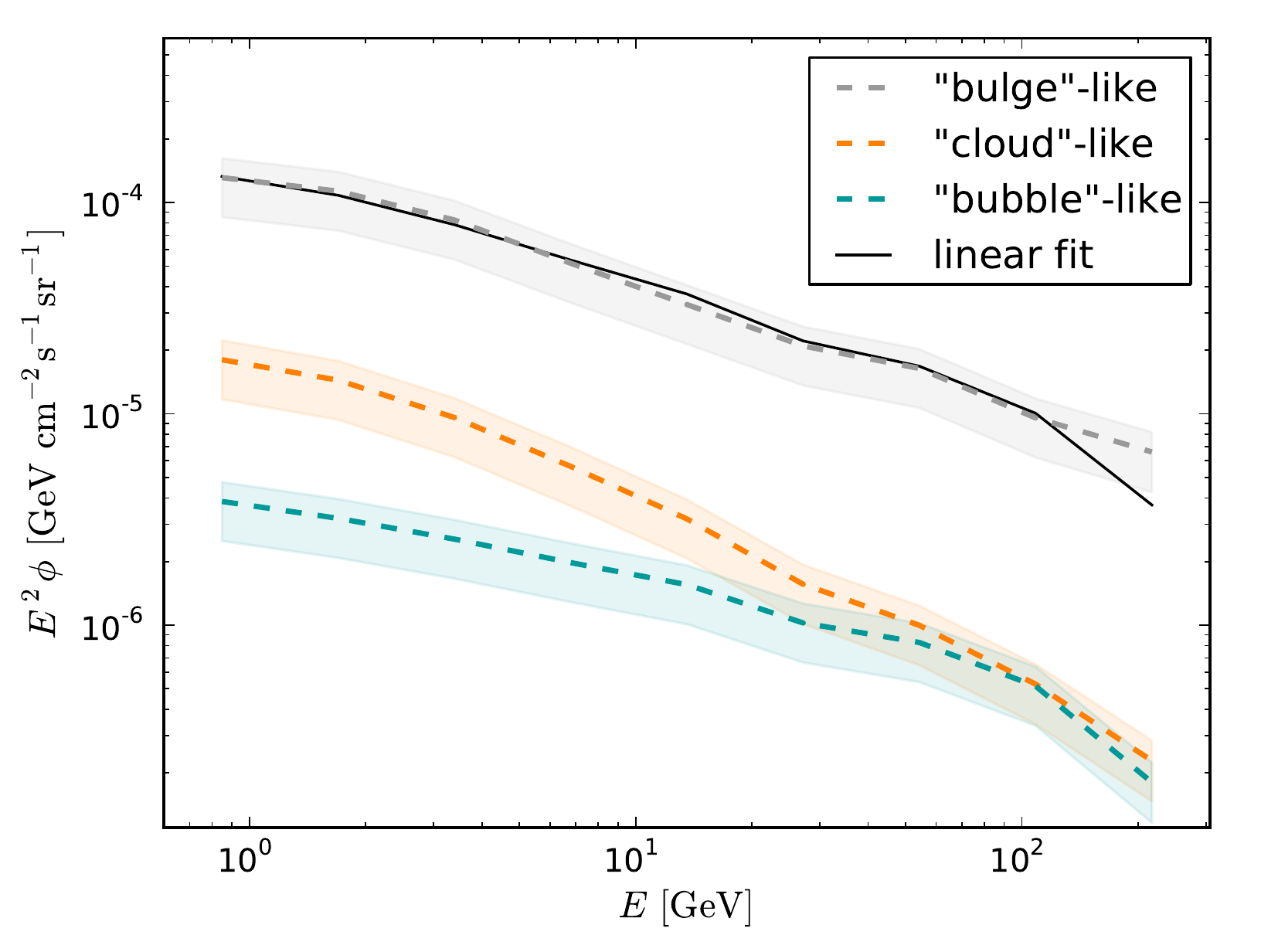} \put(0,75){(a)} \end{overpic} &
            \begin{overpic} [width=0.5\textwidth]{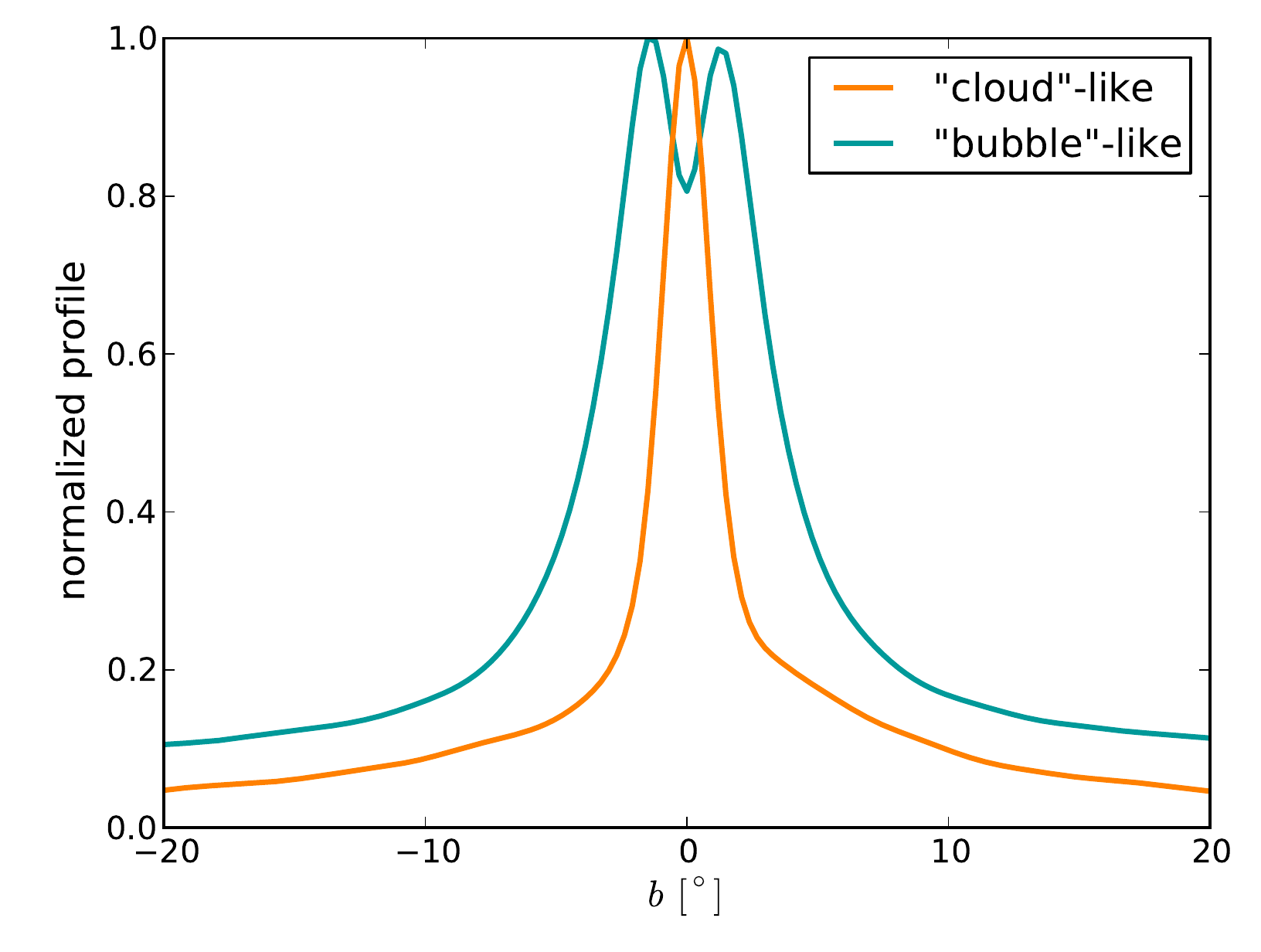}  \put(0,75){(b)} \end{overpic} \\
        \end{tabular}
        \flushleft
        \caption{Illustration of ``cloud''-, and ``bubble''-like component. Panel (a) shows energy spectra from ROIs defined in Fig.~\ref{fig:pseudo}g, cf. text. In addition to the spectra retrieved from the different regions, a linear combination of the ``cloud''-, and ``bubble''-like is fit to the ``bulge''-like component, cf. legend. Panel (b) shows the normalized latitude profiles of the ``cloud''-, and ``bubble''-like component.}
        \label{fig:speckpart}
    \end{figure*}

    \begin{figure*}[!t]
        \centering
        \begin{tabular}{cc}
            \begin{overpic} [width=0.47\textwidth]{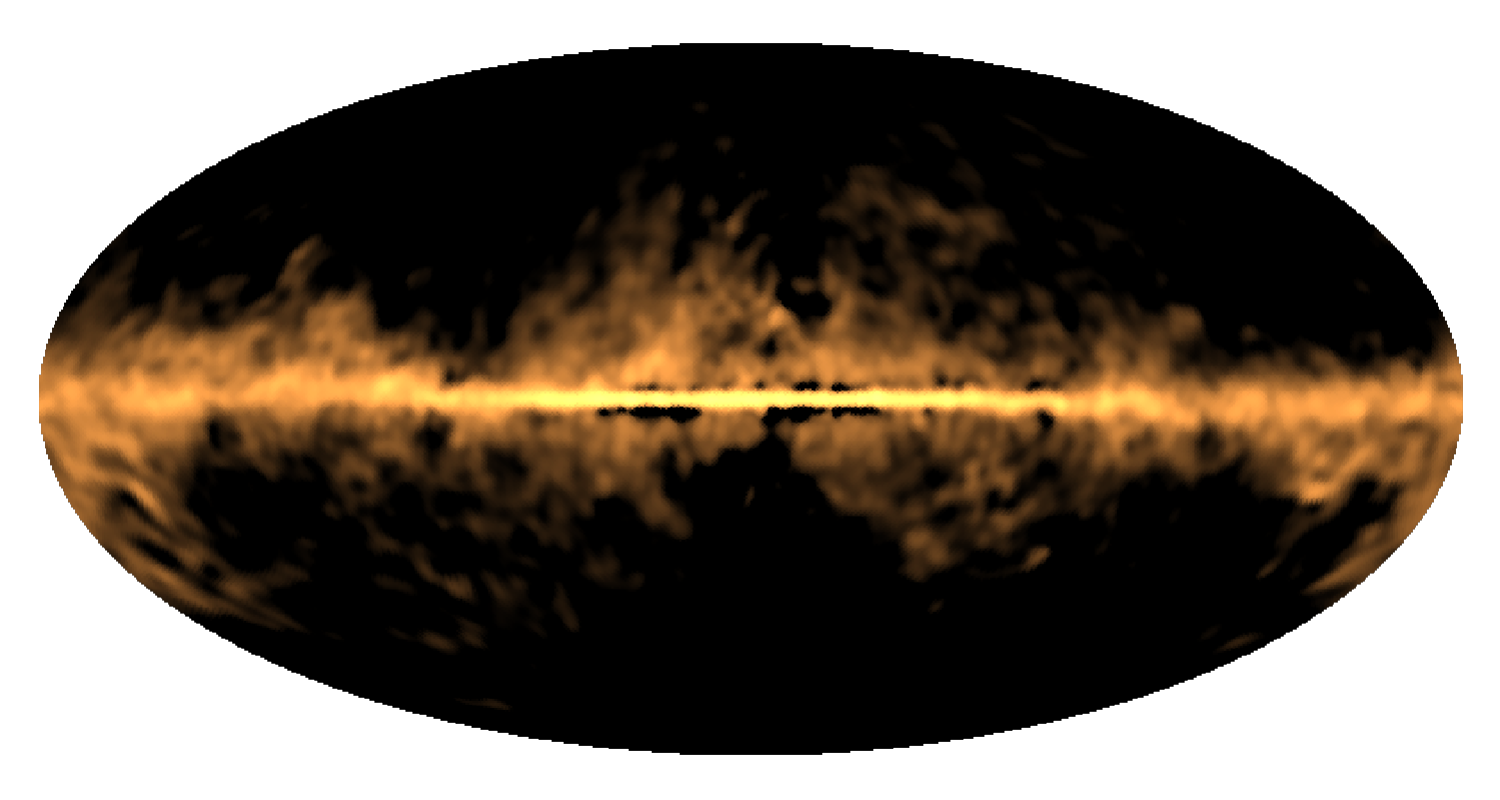}      \put(-1,51){(a)} \end{overpic} &
            \begin{overpic} [width=0.47\textwidth]{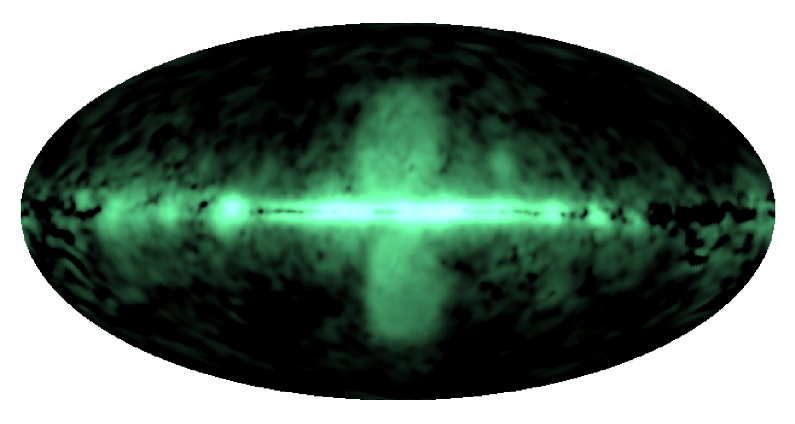}      \put(-1,51){(b)} \end{overpic} \\
        \end{tabular}
        \flushleft
        \caption{Illustration of the $\gamma$-ray sky in pseudocolor in Mollweide projection. Panel (a) shows the ``cloud''-like diffuse component and panel (b) the ``bubble''-like one.}
        \label{fig:partition}
    \end{figure*}

    \begin{figure*}[!t]
        \centering
        \begin{tabular}{ccc}
            \begin{overpic} [width=0.3\textwidth]{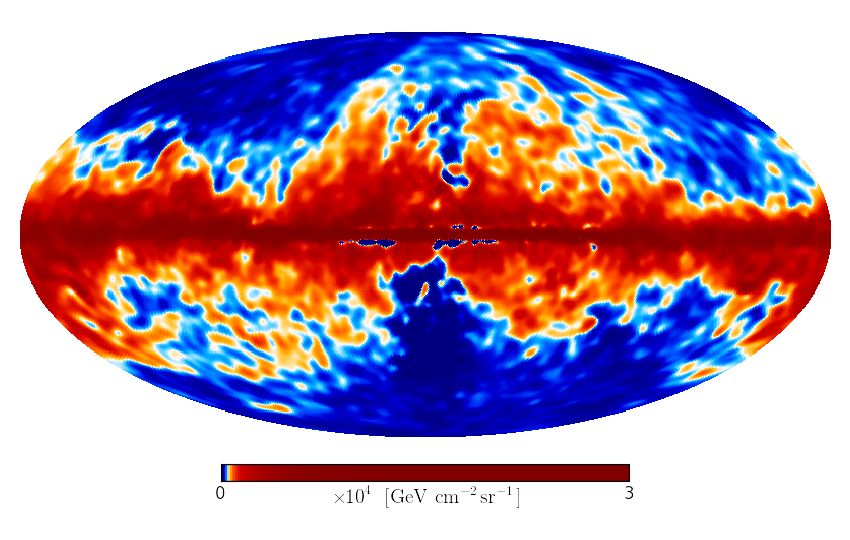}       \put(-4,60){(a)} \end{overpic} &
            \begin{overpic} [width=0.3\textwidth]{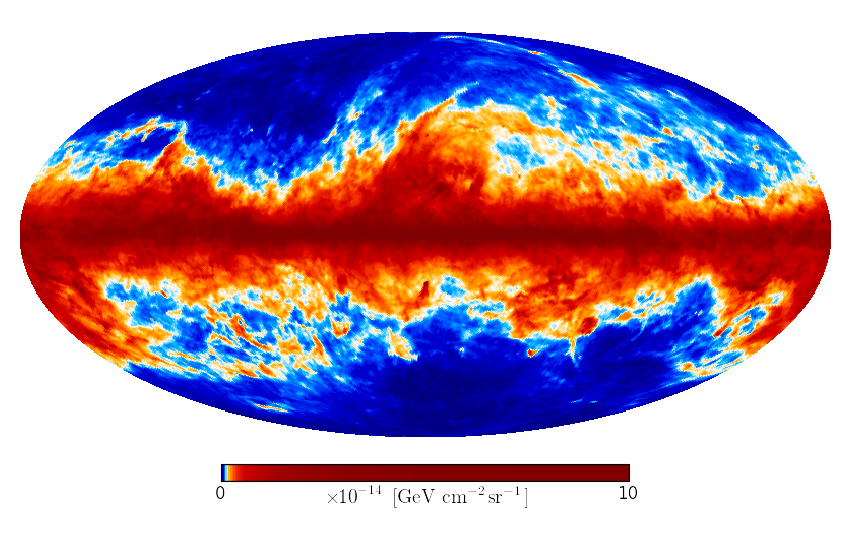}  \put(-4,60){(b)} \end{overpic} &
            \begin{overpic} [width=0.3\textwidth]{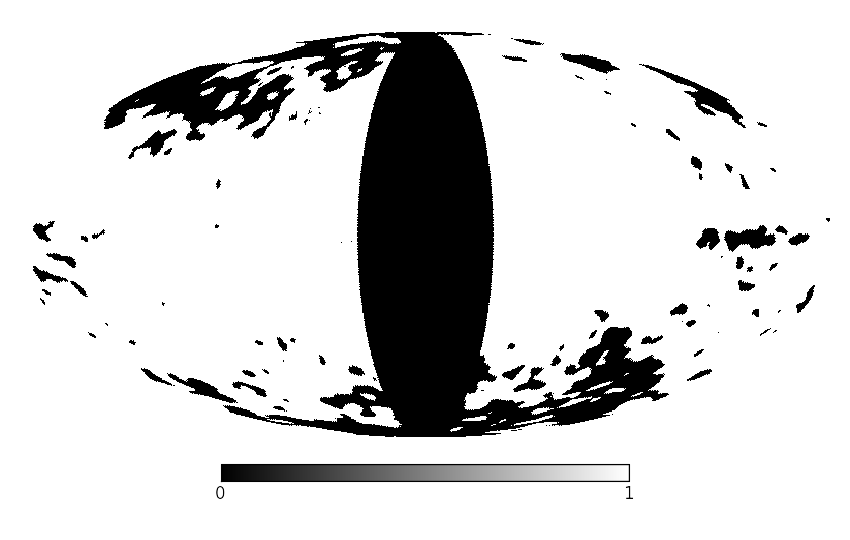} \put(-4,60){(c)} \end{overpic} \\
        \end{tabular}
        \caption{Illustration of the brightness in Mollweide view. Panel (a) shows the integrated brightness of the ``cloud''-like component. Panel (b) shows the monochromatic brightness of thermal dust emission at $353 \, \mathrm{GHz}$ \citep{P13a}. Panel (c) shows in white the area included in computing the latitude profile in Fig.~\ref{fig:profile}.}
        \label{fig:partition_int}
    \end{figure*}


    If we consider smaller ROIs, we find fluctuations in the spectral index of the diffuse $\gamma$-ray flux. These fluctuations could give some indication about the CR spectrum, the composition of the local ISM, etc.

    Since the D$^3$PO algorithm provides a continuous estimate of the diffuse photon flux, we can perform a spectral analysis in individual pixels.
    Although the energy spectra vary with location, we, for simplicity, assume a general power-law behavior everywhere, but with varying spectral index. Figure~\ref{fig:speckmap} shows the obtained spectral index map for the anisotropic $\gamma$-ray sky, centered on the average index of $2.6$.
    The spatial smoothness of the spectral index map reflects that there are no discontinuities between neighboring pixels in the reconstructions.

    From this spectral index map it is apparent that the Galactic disk is spectrally softer than the all-sky average. The same holds for the extensive structures that trace interstellar gas. These regions are dominated by hadronic interactions releasing $\gamma$-ray photons; e.g. $\pi^0$ production and decay \citep[cf{.} e.g.,][]{F12a}.

    In the region overlapping with the \emph{Giant Fermi Bubbles} we find overall similar spectra that are, however, harder than the all-sky average. This is in agreement with the results of \citet{F14a}, although they found a log-parabola to fit best.
     The strong hardening towards the high latitude edge of the southern bubble comes from its increased spatial extent compared to lower energies, cf. \citet{YAC14}.
    Further local spots inside the bubble region are insignificant within the statistical and systematic uncertainties.

    Although the morphology and spectra of the bubbles can be explained with hadronic and leptonic CR processes, IC scenarios give the preference of also reproducing the microwave haze observed with WMAP and \emph{Planck} \citep{P12a,YRZ13,F14a}.
    Furthermore, the low target densities at higher Galactic latitudes render the hadronic scenario as not very compelling.

\subsection{Diffuse emission components}
\label{sec:like}

 \begin{figure}[!b]
        \centering
        \includegraphics[width=0.5\textwidth]{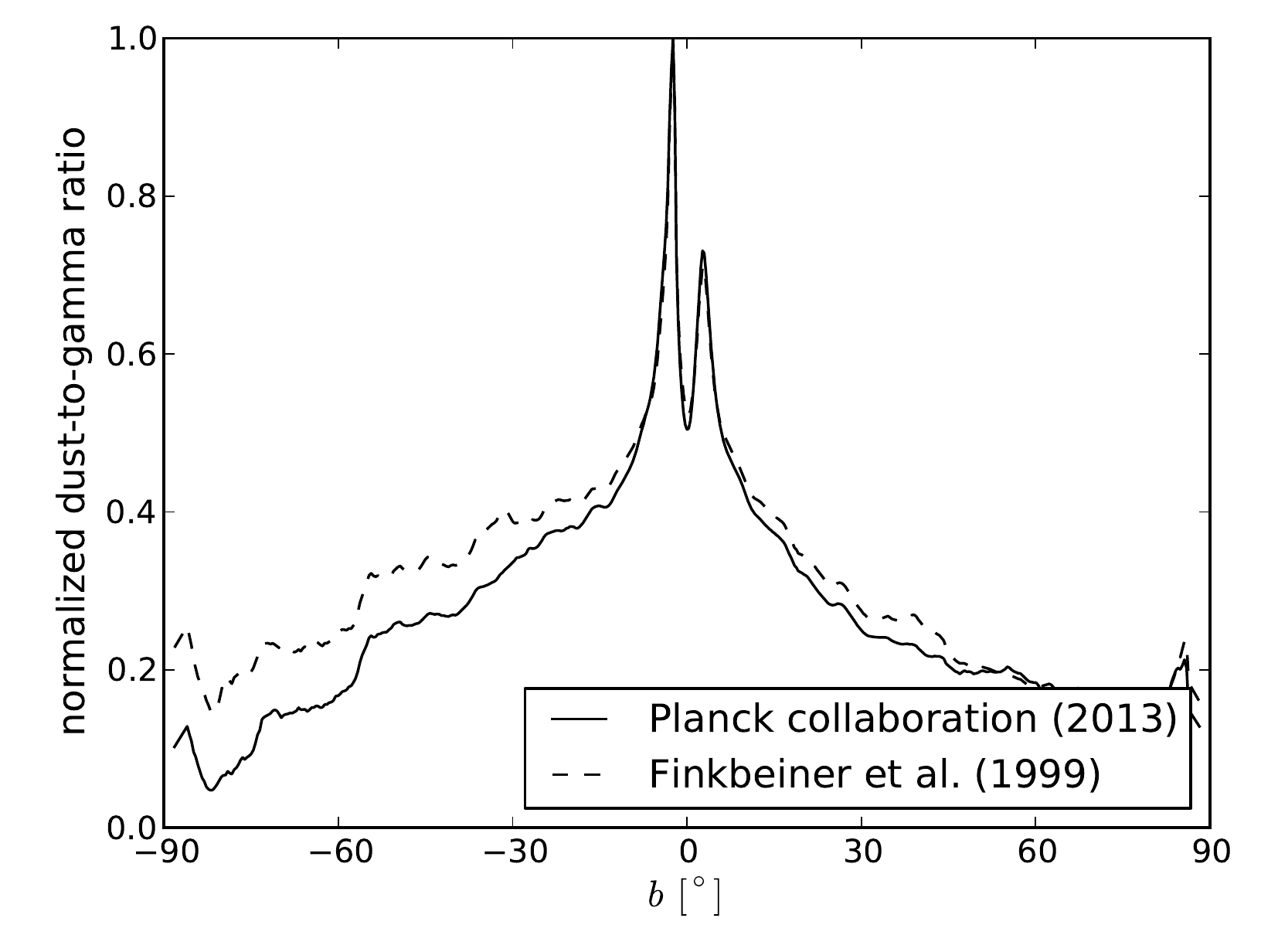}
   
        \caption{Illustration of the latitude profile of the dust-to-``cloud''-like-gamma ratio. The thermal dust emission from \citet{P13a}, and\citet{FDS99} has been smoothed with a $0.7^\circ$ Gaussian kernel to match the coarseness of the $\gamma$-ray map.}
        \label{fig:profile}
    \end{figure}
    \begin{figure}[!b]
        \centering
        \begin{tabular}{ccccc}
            \begin{overpic} [width=0.125\textwidth]{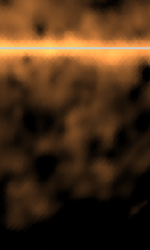} \put(-13,100){(a)} \end{overpic} \hspace{0.5em} &
            \begin{overpic} [width=0.125\textwidth]{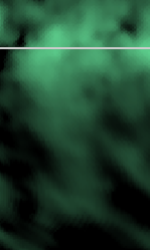} \put(-13,100){(b)} \end{overpic} \hspace{0.5em} &
            \begin{overpic} [width=0.125\textwidth]{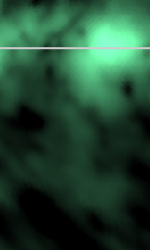} \put(-13,100){(c)} \end{overpic} \\
        \end{tabular}
  
        \caption{Panel (a) shows a magnification of the \emph{Chamaeleon} complex from Fig.~\ref{fig:partition}a. Panel (b) and (c) show magnifications of tentative outflows from Fig.~\ref{fig:partition}b. The light gray line marks the Galactic plane ($l = 0^\circ$).}
        \label{fig:zoom}
    \end{figure}

    The pseudocolor scheme introduced in Section~\ref{sec:pseudocolor} already allows us to visually inspect the continuous reconstruction of the diffuse $\gamma$-ray sky. By eye, we can make out the Galactic bulge, the \emph{Fermi} bubbles, and also cloudy structures around the Galactic anticenter.

    In order to confirm this impression the visualization gives, we retrieve energy spectra from three characteristic regions: ``bulge''-like ($|l|<40^\circ,|b|<1.5^\circ$), ``cloud''-like ($-150^\circ<l<-120^\circ,|b|<3^\circ$), and ``bubble''-like (for which we select the southern bubble up to latitudes $b<-27.5^\circ$). The contours of those regions are shown in Fig.~\ref{fig:pseudo}g.

    Figure~\ref{fig:speckpart}a shows the energy spectra retrieved from the three regions.
    The ``cloud''-like spectrum is rather soft ($\gamma^{(s)} \approx 2.6$) and features the tail of the pion bump. It is not surprising that the ``cloud''-like spectrum is dominated by emission from $\pi^0$ decay, because the cloudy structures trace the gas content of the ISM that provides the target protons for $\pi^0$ production.
    The ``bubble''-like spectrum is significantly harder ($\gamma^{(s)} \approx 2.4$) indicating the dominance of hard processes like IC emission.
    The ``bulge''-like region exhibits a spectrum that, besides having a higher absolute scale, can be described as a linear combination of the former two spectra, cf. Fig.~\ref{fig:speckpart}a.

    As the ``bulge''-like spectrum is found to be a linear combination of ``cloud''- and ``bubble''-like, we can try to decompose the whole diffuse sky into those two components. For this purpose, we fit the spectrum in individual pixels by the ``cloud''-, and ``bubble''-like component.\footnote{In case the fit suggests a negative coefficient for one component, the fitting procedure is repeated ignoring this component. In this way, we ensure the positivity of the components.}
    The fit coefficients then indicate the strength of the ``cloud''- or ``bubble''-like contribution at all locations. Multiplying the fit coefficients with the respective spectra, we obtain a pseudocolor visualization of the ``cloud''- or ``bubble''-like emission components as shown in Fig.~\ref{fig:partition}.

    In spite of the simplicity of this two-component model, we find a good agreement between the total diffuse emission and the sum of the two components.
    The relative residuals are around $5$--$13\%$, except for the highest energy band, which was excluded from the fitting procedure, where the error is approximately $28\%$.
    Our findings demonstrate that the $\gamma$-ray sky in the energy range from $0.6$ to $307.2 \, \mathrm{GeV}$ can with high precision be described by ``cloud''- and ``bubble''-like emission components only.

    From the shape of the energy spectrum of the ``cloud''-like component, we deduced that it is dominated by hadronic processes. We can also compare its morphology with other ISM tracers. For this, we compute the brightness of the ``cloud''-like component and show it in Fig.~\ref{fig:partition_int}a. The resulting map agrees with the thermal dust emission seen by \emph{Planck} \citep{P13a} at $353 \, \mathrm{GHz}$ shown in Fig.~\ref{fig:partition_int}b.
    We like to stress how similar the morphology of thermal dust microwave/IR emission and the ``cloud''-like $\gamma$-ray component are. In particular in the cloudy region in the Galactic East, the structures agree well, although we took the spectrum from a region in the West. As the thermal dust emission traces the densest part of the ISM, and hence the target gas for the CR protons, we are confident that the ``cloud''-like component is indeed dominated by hadronic emission processes.

    The \emph{Chamaeleon} complex, around $(l,b) \sim (-60^\circ,-20^\circ)$, hosting a number of star-forming clouds is visible in the ``cloud''-like component, cf. Fig.~\ref{fig:zoom}a. Recent work by the \citet{P14} used $\gamma$-, radio-, and dust data to map the local gas content of the clouds.

    The ``cloud''-like component is, however, morphologically not exactly identical to the dust emission. For example, in the latitude profile shown in Fig.~\ref{fig:profile}, the dust-to-gamma ratio decreases with increasing latitude. This profile was computed including only pixels outside the bubble region ($|l|>30^\circ$), where both, the ``cloud''- and ``bubble''-like component, contribute, and the estimated \emph{Planck} dust emission is positive.
    The dust seems to be preferentially in the Galactic disk compared to the thermal gas traced by $\gamma$-rays. 
\begin{figure*}[!t]
        \centering
        \begin{tabular}{cc}
            \begin{overpic} [width=0.5\textwidth]{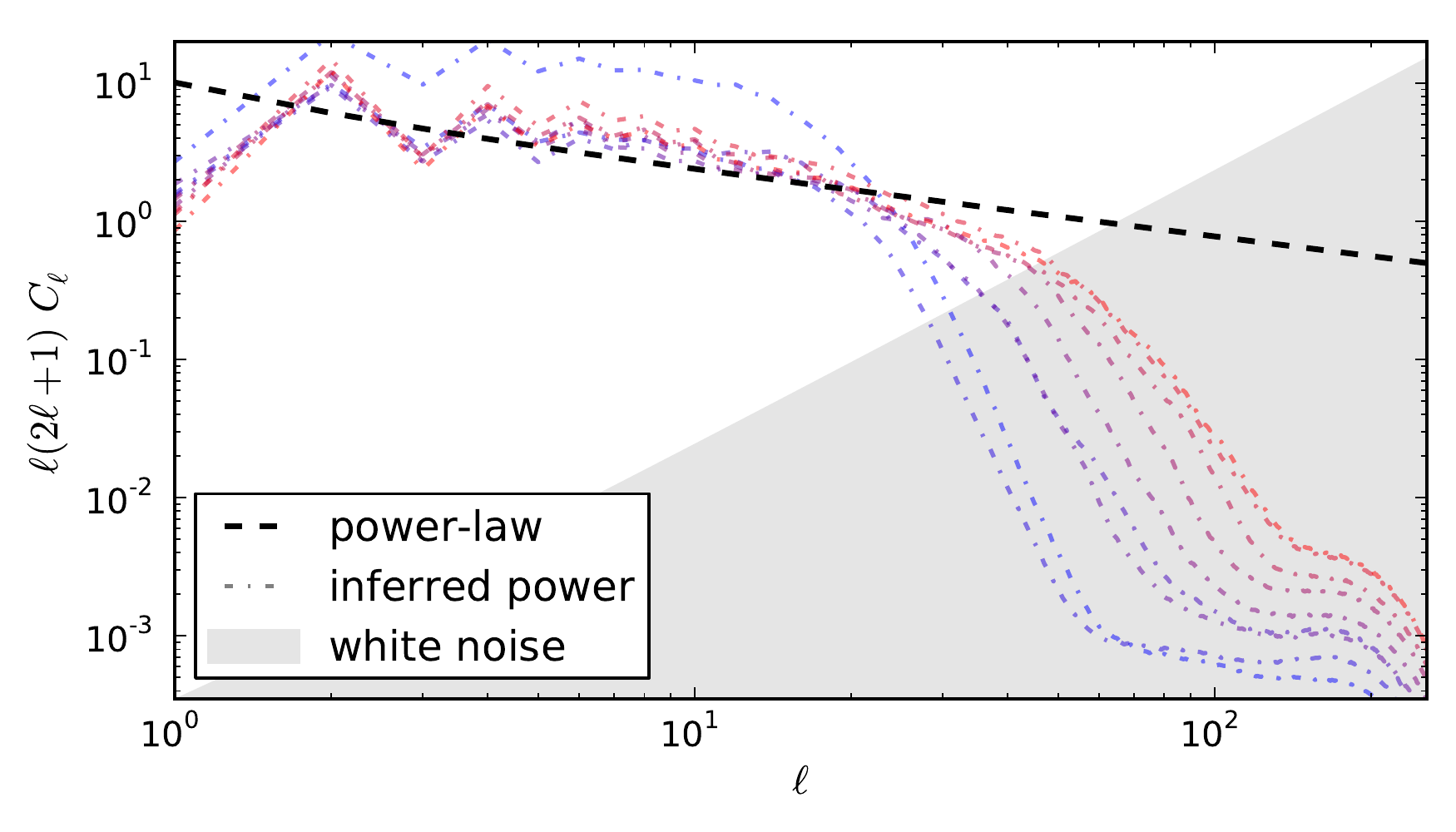} \put(0,57){(a) \hspace{1.2em} {\tiny flux power spectra}} \end{overpic} &
            \begin{overpic} [width=0.5\textwidth]{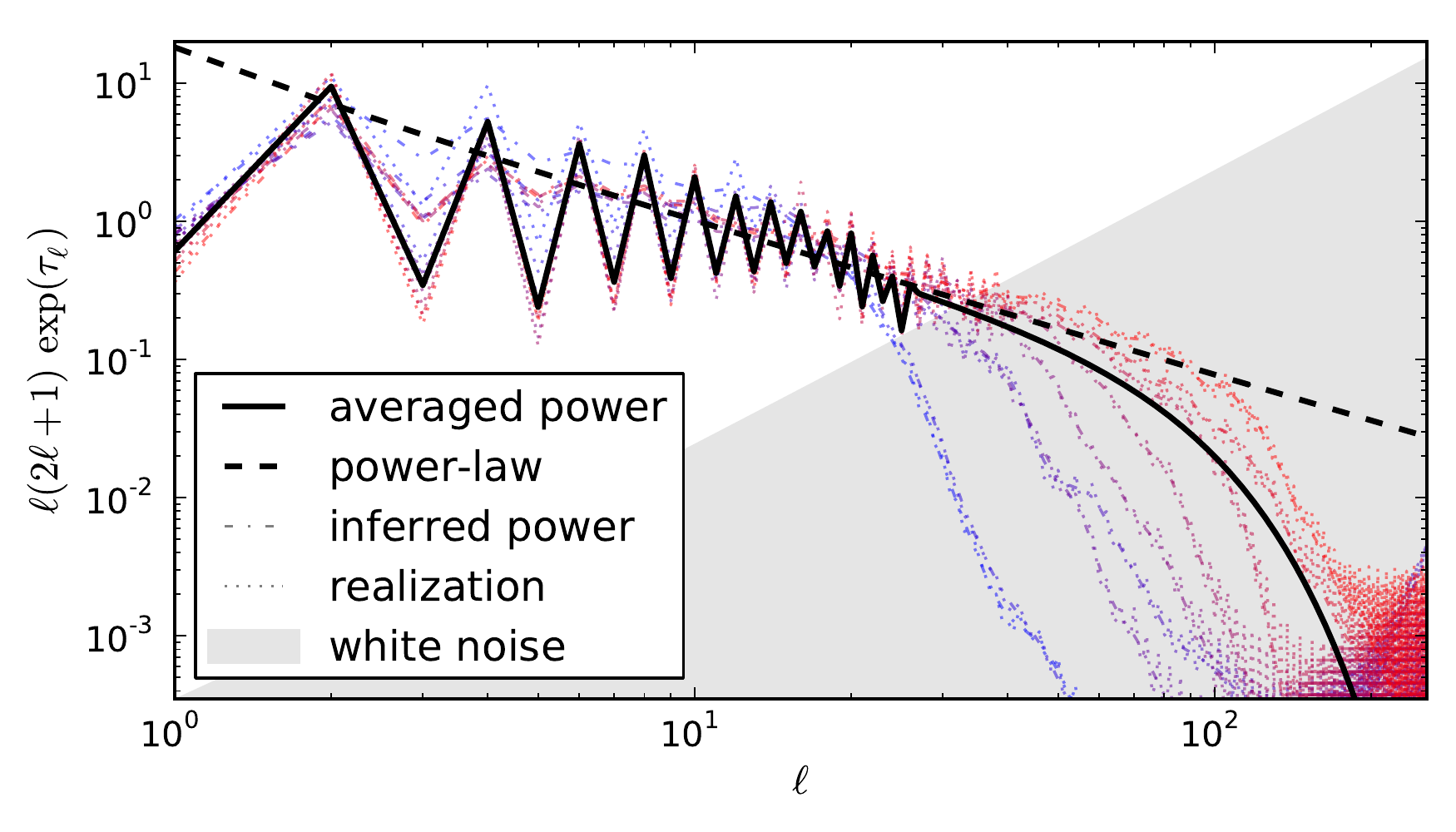} \put(0,57){(b) \hspace{1.2em} {\tiny log-flux power spectra}} \end{overpic} \\
        \end{tabular}
        \flushleft
        \caption{Illustration of angular power spectra over angular quantum number $\ell$. Panel (a) shows the inferred spectra of the diffuse photon flux $\bb{\phi}^{(s)}$ in each energy band and a power-law fit thereof, cf. legend. Panel (b) shows the direct realization spectra of the diffuse signal $s=\log(\phi^{(s)}/\phi_0)$, as well as the inferred, fitted, and averaged spectra, cf. legend. The colors (from red over violet to blue) indicate energy (from band $1$ to $9$) in both panels and a white noise spectrum with arbitrary normalization is included for comparison.}
        \label{fig:power}
    \end{figure*}
    The morphology of the ``bubble''-like component is very different, cf. Fig.~\ref{fig:partition}. Since the spectrum of the \emph{Fermi} bubbles is spatially relatively constant, the shape of the northern bubble is well recovered based on the spectrum of the southern one.
    We also find an excess in the ``bubble''-like emission in the bulge region \citep{BW14}, as well as in the star forming region \emph{Cygnus-X} around $(l,b) \sim (80^\circ,0^\circ)$.
    Excess of emission following the ``bubble''-like spectrum is also visible at intermediate latitudes exhibiting mushroom-like shapes typical for hot outflows; e.g.,  South of the \emph{Cygnus-X} region ($(l,b) \sim (90^\circ,<-15^\circ)$), or along $(l,b) \sim (130^\circ,<-15^\circ)$, cf. Fig.~\ref{fig:zoom}b and c. Those are likely candidates for outflows from active star-forming regions of the Milky Way.
    Furthermore, the latitude profile shown in Fig.~\ref{fig:speckpart}b indicates that the ``bubble''-like disk is roughly twice as thick as the ``cloud''-like component. 

    Since the ``bubble''-like $\gamma$-ray emission is morphologically so distinct, and sets itself apart from the ``cloud''-like component, we suppose that the two components are dominated by different emission processes. The ``bubble''-like spectrum is distinctly harder and less structured, therefore a leptonic emission process, in particular IC scattering, seems more convincing in causing the ``bubble''-like diffuse component.

    The CR populations producing these two $\gamma$-ray emission components do not need not to be very different.
    It might be that we are just seeing two different phases of the ISM:
    \begin{itemize}
        \item[$\bullet$]
            the cold and condensed phase carries most of the Galactic dust and has a sufficient nuclei target density to be predominately revealed through hadronic interactions with CR protons. Hence, the resulting $\gamma$-ray emission mostly traces the highly structured gas distribution.
        \item[$\bullet$]
            the hot, dilute, and voluminous phase tends to flow out of the Galactic disk. The $\gamma$-ray emission from within is dominated by IC upscattering of the Galactic photon field by CR electrons. As the photon field is relatively homogeneous, the morphology of the ``bubble''-like component is probably shaped by the spatial distribution of the CRs.
    \end{itemize}
    \noindent
    This simple two-component model of the diffuse $\gamma$-ray emission supports scenarios in which the \emph{Fermi} bubbles are just outflows of the hot ISM \citep{YRZ13,C+11,D+11,C-11,C+13}.

 \begin{figure*}[!t]
        \centering

        \includegraphics[width=\textwidth]{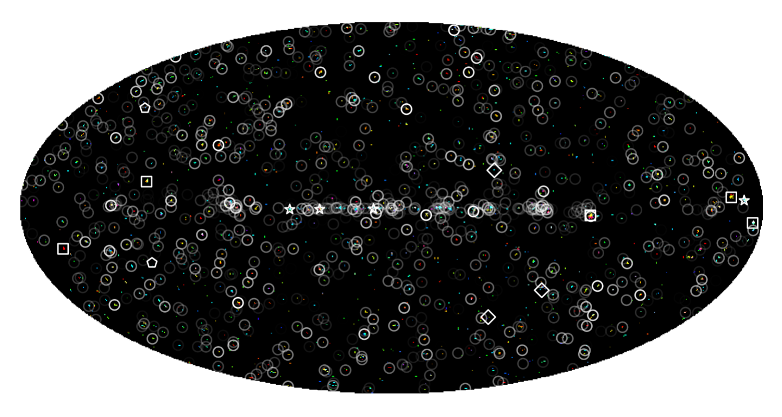}
        \flushleft
        \caption{Illustration of the point sources in the $\gamma$-ray sky in pseudocolor and in Mollweide projection. Markers show point sources from the second \emph{Fermi} LAT source catalog \citep{F12d} for comparison, whereby the gray scale indicates their average detection significance as listed in the catalog.
        Special markers show a selection of pulsars (squares), local SNRs (stars), and well-known galaxies (pentagons), as well as famous extragalactic objects (diamonds), cf. text.}
        \label{fig:sources}
    \end{figure*}

    \begin{figure*}[!t]
        \centering
        \begin{tabular}{cc}
            \begin{overpic} [width=0.5\textwidth]{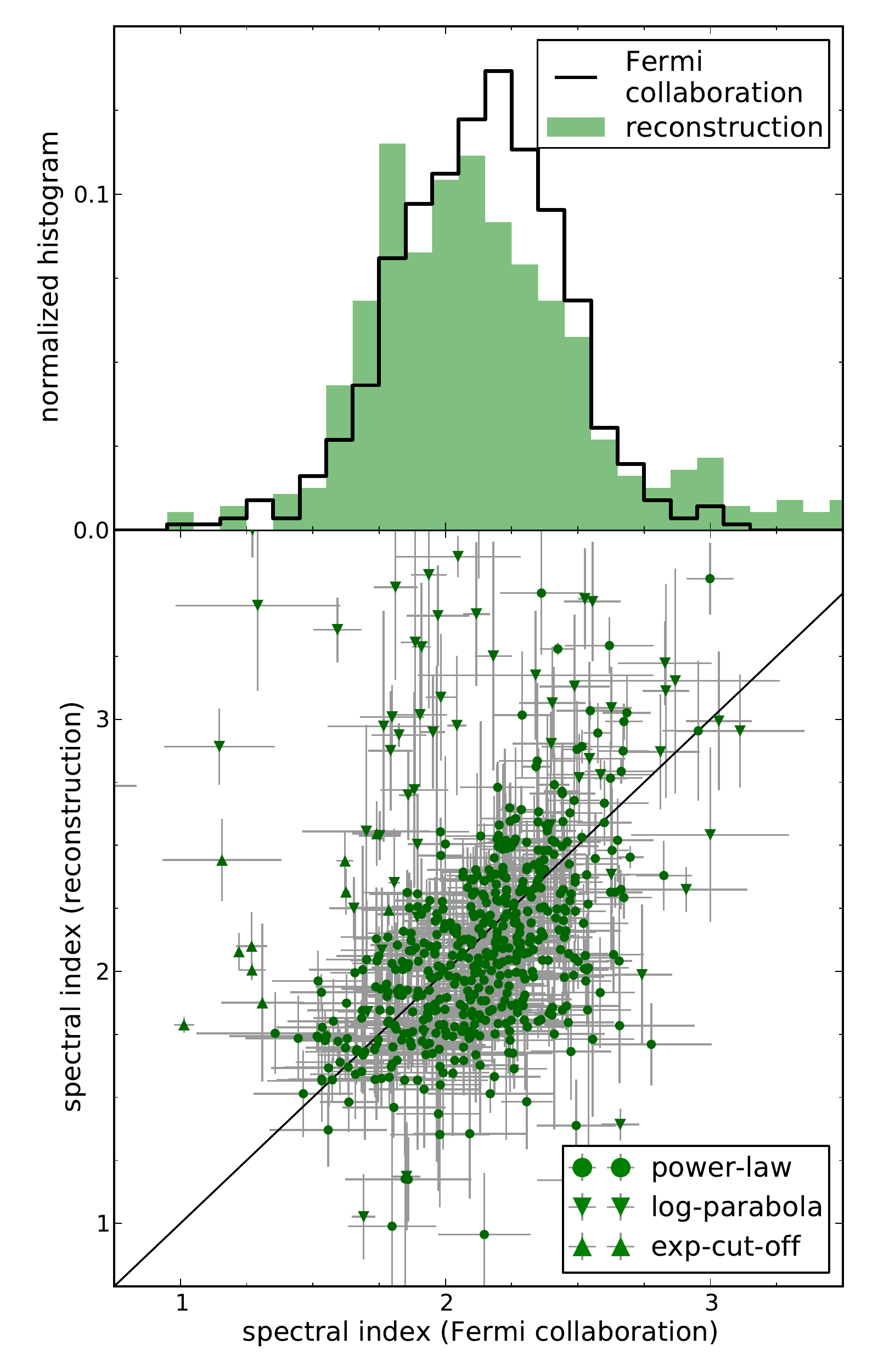} \put(0,99){(a)} \end{overpic} &
            \begin{overpic} [width=0.5\textwidth]{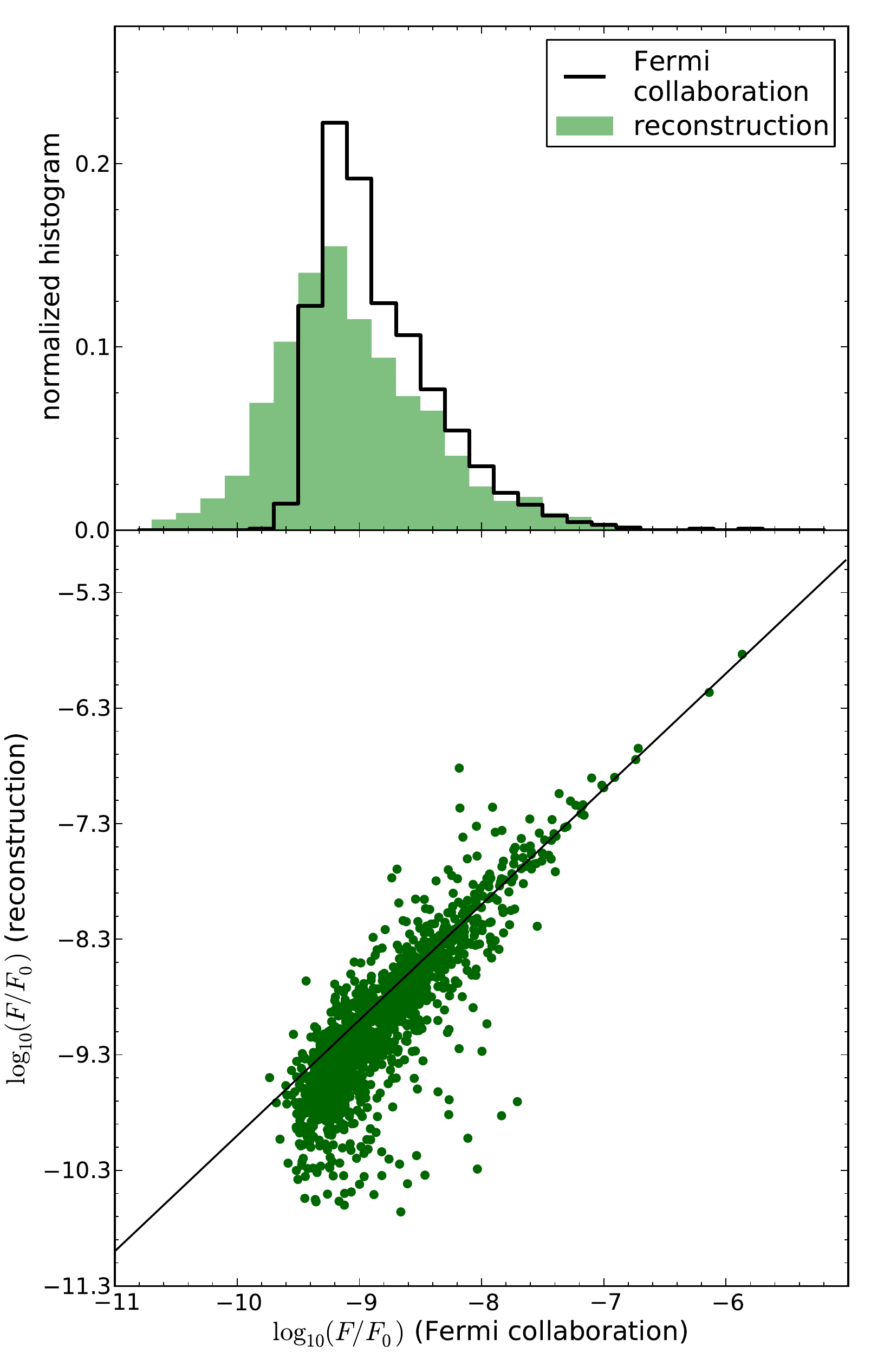}  \put(0,99){(b)} \end{overpic} \\
        \end{tabular}
        \flushleft
        \caption{Comparison of the second \emph{Fermi} LAT source catalog \citep{F12d} and candidates from the reconstruction for which an association in the second \emph{Fermi} LAT catalog has been found. Panel (a) shows the histogram and scatter plot of the spectral indices $\gamma^{(u)}$ of candidates versus catalog sources. In the latter, a $1:1$ line (black solid) is plotted for comparison, and the markers indicate the fit spectral shape, cf. legend. Panel (b) shows the histogram and scatter plot of the logarithmic total fluxes $\log_{10}(F/F_0)$, where $F_0 = 1 \, \mathrm{cm}^{-2} \, \mathrm{s}^{-1}$. The scatter plot contains a $1:1$ line (black solid) for comparison.}
        \label{fig:scatter}
    \end{figure*}
\subsection{Angular power spectra}
\label{sec:power}

    Under the assumption of statistical isotropy and homogeneity the second moments of a diffuse signal field are defined by its angular power spectrum. Studying power spectra gives some indication of the strength of typical fluctuations on respective angular scales described by the angular quantum number $\ell$. According to our chosen \textsc{HEALPix} discretization, we examine spectra up to a maximum scale set by $\ell_\mathrm{max} = 2 \, n_\mathrm{side} = 256$.

    Figure~\ref{fig:power} shows the angular power spectra of the diffuse photon flux $\bb{\phi}^{(s)}$ -- the quantity we are interested in -- and its logarithm, the diffuse signal field $\bb{s}$.
    The power spectrum, which is \emph{a~priori} unknown, needs to be reconstructed from the data alongside the diffuse signal field \citep{WLL04,JKWE10,EF11,OSBE12,J13}. This is done for each energy band separately. Further details on the inference procedure can be found in Appendix~\ref{app:ana}.

    The inferred power spectra of $\bb{s}$ show similar power-law behavior at all energies with some remarkable deviations.
    On large scales, $0 < \ell \lesssim 28$, the spectra exhibit a strong distinction between even and odd $\ell$-modes. The reason for this is the dominant contribution of the Galactic disk centered around $b = 0^\circ$ to the diffuse photon flux, which excites/suppresses even/odd $\ell$-modes in the reconstruction.
    On smaller scales, on the other hand, the power spectra start to fall off because small-scale features cannot be resolved due to the finite exposure of the \emph{Fermi} LAT. This effect has a clear energy dependence. Since events with higher energies are rarer, the decline of spectra from high-energy bands begins at lower $\ell$.
    Notice that the threshold set by the PSF is on very small scales; e.g., the $68\%$ angular containment radius above $10 \, \mathrm{GeV}$ is smaller than $0.2^\circ$ corresponding to $\ell \gtrsim 900$.

    Let us now consider the actual angular power spectrum $C_\ell$ of the diffuse photon flux $\bb{\phi}^{(s)} \propto \exp(\bb{s})$. This can be computed by transforming\footnote{Here, we disregard the respective monopole (mode with $\ell = 0$) for convenience. As a consequence, the absolute scale of the power spectrum of the diffuse $\gamma$-ray flux $\bb{\phi}^{(s)}$ becomes ambiguous.} the inferred (final) power spectra of $\bb{s}$ according to \citet{GE13}. This transformation shifts power between different $\ell$-modes, in particular toward smaller scales (larger $\ell$).
    Again, we find a power-law behavior of the angular power spectrum, as can be seen in Fig.~\ref{fig:power}a. We fit a power-law with index $\gamma_\ell^{(s)} = 2.47 \pm 0.02$. There is an energy-dependent break point, though, beyond which the power drops. The break point should shift to higher $\ell$ with increasing observation time.

\subsection{Point sources}

    Another result of our analysis of the $6.5$ year \emph{Fermi} data is a reconstruction of the point-like contribution to the photon flux, $\bb{\phi}^{(u)}$, which consists of Galactic and extragalactic point sources.

    Figure~\ref{fig:sources} shows an all-sky map of all point source candidates in the pseudocolor scheme introduced in Sec.~\ref{sec:pseudocolor}. Markers (and their opacity) in the map indicate the position (and detection significance) of point sources from the second \emph{Fermi} LAT source catalog \citep{F12d}.\footnote{For further details regarding the \emph{Fermi} LAT source catalog see \url{http://fermi.gsfc.nasa.gov/ssc/data/access/lat/2yr_catalog/}.}
    There is a diversity of sources, which is why we highlight a selection by special markers. The brightest $\gamma$-ray sources are pulsars (PSRs) like \emph{Vela} (PSR J0835-4510), \emph{Geminga} (PSR J0633+1746), and \emph{Crab} (PSR J0534+2200), but there are also pulsars that have first been detected in $\gamma$-ray; e.g., LAT PSR J0007+7303 and LAT PSR J0357+3205. The Galactic disk, and especially the bulge, is clustered with point sources including, among others, supernova remnants (SNRs) like W51C, W44, W30, and IC443. There are also many homogeneously distributed extragalactic sources, for example, the starburst galaxy \emph{Cigar} (M82) or our neighboring galaxy \emph{Andromeda} (M31). Furthermore, the core of \emph{Centaurus A} (NGC 5128) and the \emph{Small} and \emph{Large Magellanic Cloud} are visible in $\gamma$-rays.

    Deriving a catalog of source candidates from the point-like flux is difficult because a point source might, for example, appear in neighboring pixels at different energies due to different noise realizations and the aforementioned energy dependence of the signal-to-noise ratio and the PSF.
    In order to nominate a candidate, we check if the point-like contribution exceeds $2\sigma$ above the diffuse emission in at least two of the energy bands $1$--$8$, which is a simple but conservative criterion taking the diffuse reconstruction uncertainty $\sigma$ into account, cf. Fig.~\ref{fig:mapsplus}. 
    Notice that we exclude the highest energy band from our search, since the point-like flux in this band seems to be contaminated as discussed in Sec.~\ref{sec:speck}.
    We refer to the compiled point source catalog as the first D$^3$PO \emph{Fermi} (1DF) catalog of $\gamma$-ray source candidates.

    Qualitatively, the point-like flux found by D$^3$PO agrees with the second \emph{Fermi} LAT source catalog as shown in Fig.~\ref{fig:sources}. A few sources appear slightly off-center or smeared out over two or more pixels. The reason for this is that such sources are positioned between grid points of the chosen \textsc{HEALPix} grid. Notice that image pixels in Fig.~\ref{fig:sources} do not represent \textsc{HEALPix} pixels.

    We find $\mathbf{3{,}106}$ source candidates, cf. Table~\ref{tab:candidates}. For comparison, the $1$, $2$, and $4$ year \emph{Fermi} LAT source catalogs comprise $1{,}451$, $1{,}873$, and $3{,}034$ sources, respectively \citep{F10c,F12d,F15}. $1{,}381$ ($1{,}897$) of our sources can be associated with known LAT sources from the second (third) catalog as the angular distance between 1DF candidate and catalog source is less than the angular resolution of our reconstruction.
    The reason why we do not confirm all objects in the second (third) \emph{Fermi} LAT source catalog is the conservative criterion we apply.
    This still leaves $1{,}253$ ($1{,}209$) new source candidates to be confirmed by future work.

    We caution that a more detailed study is necessary to confirm or reject those candidates. The analysis of the individual sources, best done on more constrained ROIs and with higher angular resolution, is left for future work.

    The present catalog includes extended objects as, e.g., SNRs. In the second \emph{Fermi} LAT catalog 6 firm identifications, 4 associations, and 58 possible associations with SNRs are given. We detect more than $55\%$ of them. 
    In particular, 5 out of their 6 firm identifications are listed as 1DF candidates. For SNRs with fluxes $F \gtrsim 10^{-8} \,\mathrm{cm}^{-2} \mathrm{s}^{-1}$, our results are in good agreement with those of the \emph{Fermi} 
    collaboration. On the other hand, for faint sources we often recover smaller fluxes indicating the missing flux might be attributed to the diffuse emission. For a few sources we estimate a larger flux than given in the Femi LAT catalog. 
    For example, we measure $1.25 \times 10^{-7} \, \mathrm{cm}^{-2} \mathrm{s}^{-1}$ for SNR W44 in comparison to $7.96 \times 10^{-8} \, \mathrm{cm}^{-2} \mathrm{s}^{-1}$ in the \emph{Fermi} LAT catalog. To overcome this issue, a 
    modeling that takes into account the presence of discrete extended sources would be required, which is left for future work.

    \begin{table*}[!t]
        \caption{Extract from the 1DF catalog. Listed are candidate ID (derived from the \textsc{HEALPix} index), position in Galactic longitude $l$ and latitude $b$, total flux between $1$--$100 \, \mathrm{GeV}$, contributing energy bands, and distance to the associated source in the second \emph{Fermi} LAT source catalog \citep{F12d}. The full catalog including uncertainties and associations with the third \emph{Fermi} LAT source catalog is available online at \url{http://www.mpa-garching.mpg.de/ift/fermi/} as a FITS table.}
        \centering
        \tiny
        \begin{tabular}{|crrccccc|}
            \hline
            candidate ID$^{\phantom{'}}$ & $l [{}^\circ]$ & $b [{}^\circ]$ & $F^{(u)} [\:\mathrm{cm}^{-2} \, \mathrm{s}^{-1}]$ & contributing energy bands & dist. $[{}^\circ]$ & primary association & $\cdots$ \\
             \hline
            \hline
            $^{\phantom{'}}$1DF103542$^{\phantom{'}}$ & $263.5$ &   $  -2.8$ & $1.5 \times 10^{-6}$ & $\checkmark\;\checkmark\;\checkmark\;\checkmark\;\checkmark\;\checkmark\;\checkmark\;\checkmark\;\checkmark$ & $0.09$ & 2FGL J0835.3$-$4510 & $\cdots$ \\
            1DF091157 & $195.2$ &   $4.3$ & $6.9 \times 10^{-7}$ & $\checkmark\;\checkmark\;\checkmark\;\checkmark\;\checkmark\;\checkmark\phantom{\;\checkmark}\;\checkmark\phantom{\;\checkmark}$ & $0.12$ & 2FGL J0633.9$+$1746 & \\
            1DF009247 & $179.9$ &  $65.0$ & $4.1 \times 10^{-8}$ & $\checkmark\;\checkmark\;\checkmark\;\checkmark\;\checkmark\;\checkmark\;\checkmark\;\checkmark\;\checkmark$ & $0.08$ & 2FGL J1104.4$+$3812 & \\
            1DF108550 & $184.6$ &  $-5.7$ & $1.8 \times 10^{-7}$ & $\checkmark\;\checkmark\;\checkmark\;\checkmark\;\checkmark\;\checkmark\;\checkmark\;\checkmark\;\checkmark$ & $0.06$ & 2FGL J0534.5$+$2201 & \\
            1DF036698 &  $63.6$ &  $38.9$ & $1.2 \times 10^{-8}$ & $\checkmark\;\checkmark\;\checkmark\;\checkmark\;\checkmark\;\checkmark\;\checkmark\;\checkmark\;\checkmark$ & $0.08$ & 2FGL J1653.9$+$3945 & \\
            1DF159099 &  $86.1$ & $-38.3$ & $1.1 \times 10^{-7}$ & $\checkmark\;\checkmark\;\checkmark\;\checkmark\;\checkmark\;\checkmark\;\checkmark\;\phantom{\checkmark\;\checkmark}$ & $0.12$ & 2FGL J2253.9$+$1609 & \\
            1DF056702 &  $88.9$ & $ 25.0$ & $1.0 \times 10^{-7}$ & $\checkmark\;\checkmark\;\checkmark\;\checkmark\;\checkmark\;\checkmark\;\phantom{\checkmark\;\checkmark\;\checkmark}$ & $0.07$ & 2FGL J1836.2$+$5926 & \\
            1DF029553 &  $21.9$ &  $43.9$ & $1.4 \times 10^{-8}$ & $\checkmark\;\checkmark\;\checkmark\;\checkmark\;\checkmark\;\checkmark\;\checkmark\;\checkmark\;\checkmark$ & $0.04$ & 2FGL J1555.7$+$1111 & \\
            1DF080298 & $119.8$ &  $10.5$ & $6.9 \times 10^{-8}$ & $\checkmark\;\checkmark\;\checkmark\;\checkmark\;\checkmark\;\checkmark\;\checkmark\;\checkmark\phantom{\;\checkmark}$ & $0.17$ & 2FGL J0007.0$+$7303 & \\
            1DF176024 &  $17.9$ & $-52.4$ & $2.3 \times 10^{-8}$ & $\checkmark\;\checkmark\;\checkmark\;\checkmark\;\checkmark\;\checkmark\;\checkmark\;\checkmark\;\checkmark$ & $0.18$ & 2FGL J2158.8$-$3013 & \\
            $\vdots$ &&&&&&& $\ddots$ \\
            \hline
        \end{tabular}
        \normalsize
        \label{tab:candidates}
    \end{table*}

    In the following, we compare the 1DF candidates for which we find a unique association within the second \emph{Fermi} LAT source catalog \citep{F12d} by means of their spectral index and total flux. Notice that the two studies are based on different data, exposure, calibration, and analysis algorithms.\footnote{The main differences are the selection of \texttt{SOURCE} (\texttt{CLEAN}) events, the $2$ ($6.5$) years of observation, and the usage of the \texttt{P7\char`_V6} (\texttt{P7REP\char`_V15}) IRFs in the second \emph{Fermi} LAT source catalog \citep{F12d} (the candidate catalog presented here).}

    The spectral index of a source should not (or at least not strongly) be influenced by those differences. For each source, we attempt to fit three different spectral shapes: a plain power-law,
    \begin{align}
        \phi^{(u)}(E) &= K \left( \frac{E}{E_0} \right)^{-\gamma^{(u)}}
        ,
    \end{align}
    a log-parabola,
    \begin{align}
        \phi^{(u)}(E) &= K \left( \frac{E}{E_0} \right)^{-\gamma^{(u)}-\beta\log(E/E_0)}
        ,
    \end{align}
    and a power-law with exponential cut-off,
    \begin{align}
        \phi^{(u)}(E) &= K \left( \frac{E}{E_0} \right)^{-\gamma^{(u)}} \; \exp\left( -\frac{E - E_0}{E_\mathrm{cut}} \right)
        .
    \end{align}
    Here $E_0 = 1 \, \mathrm{GeV}$ serves as a reference energy, and the spectral index $\gamma^{(u)}$ is a fit parameter as are $K$, $\beta$, and $E_\mathrm{cut}$. The 1DF source catalog contains the best-fit parameters for all shapes if applicable.

    Figure~\ref{fig:scatter}a shows the comparison of the 1DF spectral indices and the ones listed in the second \emph{Fermi} LAT source catalog (corrected to $E_0$ where needed).
    The scatter of spectral indices is large, but comparable to the uncertainties. We find a rough agreement, although our distribution is broadened toward higher indices.
    Most of the outliers yielding a high (low) spectral index are modeled by a log-parabola (exponential cut-off) that has an additional degree of freedom compared to the plain power law. This implies that the versatility of point source spectra might not be covered by the considered spectral shapes.

    Figure~\ref{fig:scatter}b shows a comparison of the total fluxes $F^{(u)}$, defined as
    \begin{align}
        F^{(u)} &= \int\d \Omega \, \sum_{j=1}^8 \int_{{\widetilde E}_j^\mathrm{min}}^{{\widetilde E}_j^\mathrm{max}}\d E \left(\frac{E}{E_j^\mathrm{mid}}\right)^{-\gamma^{(u)}} \phi^{(u)}(E_j^\mathrm{mid})
    \end{align}
    with $j$ labeling the energy bands and
    \begin{align}
        {\widetilde E}_j^\mathrm{max} &= \mathrm{min}\{1 \, \mathrm{GeV},\, E_j^\mathrm{min}\}
        \notag \\
        {\widetilde E}_j^\mathrm{min} &= \mathrm{max}\{E_j^\mathrm{max},\, 100 \, \mathrm{GeV}\}
        , \notag
    \end{align}
    in a histogram and a scatter plot.
    The fluxes show a good overall agreement.
    At the faint end, our fluxes tend to come below the fluxes reported in the second \emph{Fermi} LAT source catalog \citep{F12d}. Since our analysis benefits from a higher exposure and improved calibration, the fluxes from the \emph{Fermi} collaboration might rather be considered as upper limits in this comparison.

\subsubsection{Galaxy clusters}

    Some galaxy clusters exhibit diffuse, extended radio emission, so-called radio halos, which proves the existence of relativistic electrons therein. If relativistic protons are present as well, $\gamma$-ray emission is to be expected due to hadronic interactions \citep[][and references therein]{F10d,AGC2014}.

    \citet{FGGM12} provide a collection of clusters hosting radio halos. We investigate the presence of $\gamma$-ray emission in the direction of those clusters and in the direction of the clusters listed by \cite{F10d,AGC2014}, which would in our reconstruction appear point-like due to the pixelization of our reconstruction.
    Table~\ref{tab:cluster} lists our upper limits on the total flux $F^\mathrm{up}$ and the level of the diffuse emission $F^{(s)}$ at the cluster locations.
    The upper limit flux is computed according to $F^\mathrm{up} = F^{(u)} + 2 \sigma_{F^{(s)}}$, where $\sigma_{F^{(s)}}$ is the uncertainty of the total diffuse flux $F^{(s)}$.\footnote{Unfortunately, D$^3$PO does not converge on an all-sky point-like uncertainty map. When convergence is not achieved, a relative error of 1. is reported, see the uncertainty files available online.} This is the largest possible flux hidden under the diffuse $\gamma$-ray emission.

    We find upper limit fluxes between $10^{-11} \mathrm{cm}^{-2} \, \mathrm{s}^{-1}$ and $10^{-8} \mathrm{cm}^{-2} \, \mathrm{s}^{-1}$ for the energy range $1$--$100 \, \mathrm{GeV}$. 
    \citet{F10d,AGC2014} provide upper limits for nearby clusters above $0.1 \, \mathrm{GeV}$. Some clusters are in both samples, e.g., A2256, A2319, \emph{Coma} (A1656), \emph{Ophiuchus}, \emph{Perseus} (A0426), A1914, A2029, A2142, A2163, A2744, \emph{Bullet} (1E 0657-56), MACSJ0717.5+3745. We find comparable or slightly lower upper limit fluxes. For example, for the \emph{Coma} (A1656) cluster \citet{AGC2014} report about $1.13 \times 10^{-10} \mathrm{cm}^{-2} \, \mathrm{s}^{-1}$  above $1 \, \mathrm{GeV}$, which is comparable to the $1.1 \times 10^{-10} \mathrm{cm}^{-2} \, \mathrm{s}^{-1}$ we obtain. 
    However, we caution that our upper limits are not strict $95\%$ confidence intervals, as they are not a direct outcome of our inference, but estimated as described above. A closer investigation is left for future work.

    At the location of a few clusters in our sample we found a source in our point-source catalog.
    We marked these clusters in bold in Table~\ref{tab:cluster} and reported the flux instead of an upper limit. 
    Some of them are in projection to known $\gamma$-ray point-sources (active galaxies). These are, e.g., \emph{Perseus} (NGC 1275 and IC310), A2390, and Virgo. Others are already present either in the second or in the third \emph{Fermi} LAT source catalog, while ten are new detections (clusters marked with the symbol $^{\dag}$). 
    Noteworthy among the new detections are Hydra-A, and seven clusters hosting large-scale diffuse synchrotron emission. These are: A209, A2254, and RXCJ1514.9-1523 (radio halo), A2029, A2626, and RXJ1347.5-1145 (radio mini-halo), and A4038 (radio relic). An analysis at higher resolution is necessary to localize and better understand the origin of this emission.

    \begin{table}[!t]
        \caption{Overview of total flux upper limits for clusters hosting a radio (mini-)halo from the collection by \citet{FGGM12}. Clusters reported by \cite{F10d,AGC2014} are also listed. For $\gamma$-ray detected clusters
we give the actual flux estimates and mark those in bold. The symbol $^{\dag}$ marks clusters present in our point source catalog for which an association with known sources is not yet available.}
        \centering
        \tiny
        \begin{tabular}{|lcc|}
            \hline
            cluster name$^{\phantom{'}}$ & $F_{\mathrm{up}/\mathbf{obs}} [\:\mathrm{cm}^{-2} \, \mathrm{s}^{-1}]$ & $F^{(s)} [\:\mathrm{cm}^{-2} \, \mathrm{s}^{-1}]$ \\
            \hline
            \hline
{\bf A85} &  $ \mathbf{8.5 \times  10^{-11} }$ & $ 8.4 \times 10^{-11} $\\
A119 & $ 9.0 \times 10^{-11} $ & $ 8.2 \times 10^{-11 } $\\
A133 & $ 5.4 \times 10^{-11} $ & $ 6.2 \times 10^{-11 } $\\
{\bf A209}$^{\dag}$ & $ \mathbf{7.6 \times 10^{-11} }$ & $ 6.3 \times 10^{-11 } $\\
A262 & $ 9.8 \times 10^{-11} $ & $ 1.2 \times 10^{-10 } $\\
A399 & $ 8.7 \times 10^{-11} $ & $ 1.6 \times 10^{-10 } $\\
{\bf A400} & $ 4.2 \times 10^{-10} $ & $ 1.7 \times 10^{-10 } $\\
A401 & $ 1.2 \times 10^{-10} $ & $ 1.6 \times 10^{-10 } $\\
{\bf A426} & $ \mathbf{3.1 \times 10^{-08}} $ & $ 2.6 \times 10^{-10 } $\\
{\bf A478}$^{\dag}$ & $ \mathbf{3.1 \times 10^{-10}} $ & $ 2.8 \times 10^{-10 } $\\
A496 & $ 7.6 \times 10^{-11} $ & $ 1.2 \times 10^{-10 } $\\
A520 & $ 8.6 \times 10^{-11} $ & $ 1.2 \times 10^{-10 } $\\
A521 & $ 9.1 \times 10^{-11} $ & $ 1.2 \times 10^{-10 } $\\
A523 & $ 9.1 \times 10^{-11} $ & $ 1.8 \times 10^{-10 } $\\
A545 & $ 2.1 \times 10^{-10} $ & $ 1.9 \times 10^{-10 } $\\
A548e & $ 5.9 \times 10^{-11} $ & $ 7.9 \times 10^{-11 } $\\
A576 & $ 1.0 \times 10^{-10} $ & $ 1.2 \times 10^{-10 } $\\
A665 & $ 5.7 \times 10^{-11} $ & $ 1.1 \times 10^{-10 } $\\
A697 & $ 5.4 \times 10^{-11} $ & $ 8.8 \times 10^{-11 } $\\
A746 & $ 4.5 \times 10^{-11} $ & $ 7.5 \times 10^{-11 } $\\
A754 & $ 1.5 \times 10^{-10} $ & $ 1.0 \times 10^{-10 } $\\
A773 & $ 7.7 \times 10^{-11} $ & $ 7.5 \times 10^{-11 } $\\
A781 & $ 5.7 \times 10^{-11} $ & $ 7.8 \times 10^{-11 } $\\
A851 & $ 4.7 \times 10^{-11} $ & $ 7.1 \times 10^{-11 } $\\
A1060 & $ 1.4 \times 10^{-10} $ & $ 1.3 \times 10^{-10 } $\\
A1213 & $ 7.6 \times 10^{-11} $ & $ 7.7 \times 10^{-11 } $\\
A1300 & $ 2.2 \times 10^{-10} $ & $ 1.3 \times 10^{-10 } $\\
A1351 & $ 5.1 \times 10^{-11} $ & $ 7.3 \times 10^{-11 } $\\
A1367 & $ 6.7 \times 10^{-11} $ & $ 8.6 \times 10^{-11 } $\\
A1644 & $ 1.0 \times 10^{-10} $ & $ 1.4 \times 10^{-10 } $\\
A1656 & $ 1.1 \times 10^{-10} $ & $ 7.2 \times 10^{-11 } $\\
A1689 & $ 8.2 \times 10^{-11} $ & $ 9.8 \times 10^{-11 } $\\
A1758a & $ 1.2 \times 10^{-10} $ & $ 6.9 \times 10^{-11 } $\\
A1795 & $ 7.6 \times 10^{-11} $ & $ 6.7 \times 10^{-11 } $\\
A1835 & $ 6.3 \times 10^{-11} $ & $ 1.0 \times 10^{-10 } $\\
A1914 & $ 1.1 \times 10^{-10} $ & $ 8.1 \times 10^{-11 } $\\
A1995 & $ 5.7 \times 10^{-11} $ & $ 6.8 \times 10^{-11 } $\\
{\bf A2029}$^{\dag}$ & $\mathbf{ 2.2 \times 10^{-10}} $ & $ 1.5 \times 10^{-10 } $\\
A2034 & $ 5.2 \times 10^{-11} $ & $ 7.5 \times 10^{-11 } $\\
A2065 & $ 9.0 \times 10^{-11} $ & $ 1.0 \times 10^{-10 } $\\
A2142 & $ 6.2 \times 10^{-11} $ & $ 1.1 \times 10^{-10 } $\\
A2163 & $ 2.6 \times 10^{-10} $ & $ 3.0 \times 10^{-10 } $\\
A2199 & $ 7.3 \times 10^{-11} $ & $ 9.2 \times 10^{-11 } $\\
A2218 & $ 5.9 \times 10^{-11} $ & $ 8.6 \times 10^{-11 } $\\
A2219 & $ 1.1 \times 10^{-10} $ & $ 8.2 \times 10^{-11 } $\\
A2244 & $ 6.0 \times 10^{-11} $ & $ 8.4 \times 10^{-11 } $\\
{\bf A2254}$^{\dag}$ & $ \mathbf{7.2 \times 10^{-11} }$ & $ 1.5 \times 10^{-10 } $\\
A2255 & $ 1.2 \times 10^{-10} $ & $ 8.9 \times 10^{-11 } $\\
A2256 & $ 6.8 \times 10^{-11} $ & $ 1.0 \times 10^{-10 } $\\
A2294 & $ 8.6 \times 10^{-11} $ & $ 1.5 \times 10^{-10 } $\\
A2319 & $ 1.3 \times 10^{-10} $ & $ 1.7 \times 10^{-10 } $\\
{\bf A2390} & $ \mathbf{5.3 \times 10^{-11}} $ & $ 1.6 \times 10^{-10 } $\\
A2589 & $ 1.1 \times 10^{-10} $ & $ 1.1 \times 10^{-10 } $\\
A2597 & $ 5.1 \times 10^{-11} $ & $ 6.6 \times 10^{-11 } $\\
{\bf A2626}$^{\dag}$ & $ \mathbf{7.2 \times 10^{-11}} $ & $ 1.2 \times 10^{-10 } $\\
A2634 & $ 7.6 \times 10^{-11} $ & $ 1.1 \times 10^{-10 } $\\
A2657 & $ 1.1 \times 10^{-10} $ & $ 1.2 \times 10^{-10 } $\\
A2734 & $ 5.7 \times 10^{-11} $ & $ 6.3 \times 10^{-11 } $\\
A2744 & $ 8.2 \times 10^{-11} $ & $ 6.3 \times 10^{-11 } $\\
\hline
        \end{tabular}
        \normalsize
        \label{tab:cluster}
    \end{table}
    \begin{table}[!t]
        \centering
        \tiny
        \begin{tabular}{|lcc|}
            \hline
            cluster name$^{\phantom{'}}$ & $F_{\mathrm{up}/\mathbf{obs}} [\:\mathrm{cm}^{-2} \, \mathrm{s}^{-1}]$ & $F^{(s)} [\:\mathrm{cm}^{-2} \, \mathrm{s}^{-1}]$ \\
            \hline
            \hline
A2877 & $ 4.2 \times 10^{-11} $ & $ 5.6 \times 10^{-11 } $\\
A3112 & $ 4.1 \times 10^{-11} $ & $ 6.2 \times 10^{-11 } $\\
A3158 & $ 5.2 \times 10^{-11} $ & $ 6.6 \times 10^{-11 } $\\            
A3266 & $ 8.9 \times 10^{-11} $ & $ 8.3 \times 10^{-11 } $\\
A3376  & $2.3\times 10^{-10}$ & $1.3\times 10^{-10}$ \\
A3526 & $ 1.5 \times 10^{-10 } $ & $ 2.1 \times 10^{-10}$\\
A3562 & $ 9.5 \times 10^{-11 } $ & $ 1.6 \times 10^{-10 }$\\
A3571 & $ 1.1 \times 10^{-10 } $ & $ 1.7 \times 10^{-10 }$\\
A3627 & $ 2.1 \times 10^{-10 } $ & $ 4.7 \times 10^{-10 }$\\
A3822 & $ 8.9 \times 10^{-11 } $ & $ 8.9 \times 10^{-11 }$\\
A3827 & $ 6.4 \times 10^{-11 } $ & $ 8.4 \times 10^{-11 }$\\
{\bf A3921}$^{\dag}$ & $\mathbf{ 1.2 \times 10^{-11 } }$ & $ 7.7 \times 10^{-11 }$\\
{\bf A4038}$^{\dag}$ & $ \mathbf{1.6 \times 10^{-11 }} $ & $ 6.8 \times 10^{-11 }$\\
A4059 & $ 8.5 \times 10^{-11 } $ & $ 5.6 \times 10^{-11 }$\\
1E0657-56 & $ 7.6 \times 10^{-11 } $ & $ 1.5 \times 10^{-10 }$\\
2A0335+096 & $ 1.9 \times 10^{-10 } $ & $ 2.7 \times 10^{-10 }$\\
3C129 & $ 3.7 \times 10^{-10 } $ & $ 8.6 \times 10^{-10 }$\\
AWM7 & $ 1.2 \times 10^{-10 } $ & $ 1.7 \times 10^{-10 }$\\
CIZAJ2242.8+5301 & $ 2.8 \times 10^{-10 } $ & $ 3.7 \times 10^{-10 }$\\
CL0016+16 & $ 6.5 \times 10^{-11 } $ & $ 1.1 \times 10^{-10 }$\\
CL0217+70 & $ 2.6 \times 10^{-10 } $ & $ 5.7 \times 10^{-10 }$\\
EXO0422-086 & $ 1.4 \times 10^{-10 } $ & $ 1.3 \times 10^{-10 }$\\
FORNAX & $ 6.7 \times 10^{-11 } $ & $ 6.8 \times 10^{-11 }$\\
HCG94 & $ 1.6 \times 10^{-10 } $ & $ 1.4 \times 10^{-10 }$\\
{\bf HYDRA-A}$^{\dag}$ & $\mathbf{ 3.4 \times 10^{-11 } }$ & $ 9.5 \times 10^{-11 }$\\
IIIZw54 & $ 1.1 \times 10^{-10 } $ & $ 2.3 \times 10^{-10 }$\\
IIZw108 & $ 8.1 \times 10^{-11 } $ & $ 1.2 \times 10^{-10 }$\\
M49 & $ 6.0 \times 10^{-11 } $ & $ 9.2 \times 10^{-11 }$\\
MACSJ0717.5+3745 & $ 1.2 \times 10^{-10 } $ & $ 1.2 \times 10^{-10 }$\\
MRC0116+111 & $ 7.6 \times 10^{-11 } $ & $ 9.6 \times 10^{-11 }$\\
NGC1550 & $ 1.0 \times 10^{-10 } $ & $ 2.0 \times 10^{-10 }$\\
NGC5044 & $ 1.1 \times 10^{-10 } $ & $ 1.4 \times 10^{-10 }$\\
NGC4636 & $ 7.3 \times 10^{-11 } $ & $ 1.0 \times 10^{-10 }$\\
NGC5813 & $ 1.2 \times 10^{-10 } $ & $ 1.5 \times 10^{-10 }$\\
NGC5846 & $ 1.4 \times 10^{-10 } $ & $ 1.7 \times 10^{-10 }$\\
Ophiuchus & $ 2.9 \times 10^{-10 } $ & $ 7.0 \times 10^{-10 }$\\
{\bf RBS797} & $ \mathbf{ 6.5 \times 10^{-11} }$ & $ 7.5 \times 10^{-11  }$\\
RXJ0107.7+5408 & $ 1.5 \times 10^{-10 } $ & $ 3.4 \times 10^{-10 }$\\
RXCJ1314.4-2515 & $ 9.6 \times 10^{-11 } $ & $ 1.9 \times 10^{-10 }$\\
{\bf RXCJ1514.9-1523}$^{\dag}$ & $ \mathbf{ 4.0 \times 10^{-11 } }$ & $ 2.0 \times 10^{-10 }$\\
RXCJ2003.5-2323 & $ 3.5 \times 10^{-10 } $ & $ 2.1 \times 10^{-10 }$\\
{\bf RXJ1347.5-1145}$^{\dag}$ & $\mathbf{ 5.5 \times 10^{-11}} $ & $ 1.5 \times 10^{-10 }$\\
RXCJ1504.1-0248 & $ 1.9 \times 10^{-10 } $ & $ 2.0 \times 10^{-10 }$\\
RXCJ2344.2-0422 & $ 7.1 \times 10^{-11 } $ & $ 9.0 \times 10^{-11 }$\\
S405 & $ 6.9 \times 10^{-11 } $ & $ 1.3 \times 10^{-10 }$\\
S540 & $ 5.7 \times 10^{-11 } $ & $ 9.9 \times 10^{-11 }$\\
S636 & $ 1.2 \times 10^{-10 } $ & $ 1.5 \times 10^{-10 }$\\
TRIANGULUM & $ 1.2 \times 10^{-10 } $ & $ 2.9 \times 10^{-10 }$\\
UGC03957 & $ 2.0 \times 10^{-10 } $ & $ 1.2 \times 10^{-10 }$\\
{\bf VIRGO} & $\mathbf{ 1.2 \times 10^{-09 } }$ & $ 1.1 \times 10^{-10 }$\\
ZwCl1742.1+3306 & $ 1.5 \times 10^{-09} $ & $ 1.2 \times 10^{-10 }$\\

            \hline
        \end{tabular}
        \normalsize
    \end{table}

\section{Conclusions and summary}
\label{sec:conclusion}

    We analyze the \emph{Fermi} LAT $6.5$ year photon data in the energy range from $0.6$ to $307.2 \, \mathrm{GeV}$. Applying the D$^3$PO inference algorithm, the data are effectively denoised, deconvolved, and decomposed with Bayesian inference methods. In contrast to previous approaches by the \emph{Fermi} collaboration and others, our non-parametric reconstruction does not rely on emission templates.

    We obtain estimates for the diffuse and point-like contributions to the $\gamma$-ray flux. Furthermore, D$^3$PO also provides uncertainty information and the \emph{a~priori} unknown angular power spectrum of the diffuse contribution.

    The inferred diffuse photon flux reveals the diversity of the $\gamma$-ray sky. We clearly reproduce the structure of the Galactic disk, bulge, and local interstellar gas, all of which are primarily illuminated by photons induced by hadronic interactions of CRs with the ISM.
    We confirm the existence of the \emph{Giant Fermi Bubbles}, as well as their homogeneous morphology, sharp edges, and hard spectra.
    Moreover, we are also able to resolve small diffuse structures; e.g., the $\gamma$-ray glow around \emph{Centaurus A}.

    The continuous reconstruction of the diffuse flux allows us to present the first spectral index map of the diffuse $\gamma$-ray sky, as well as a pseudocolor composite visualizing the spectrally different regions.
    Furthermore, the large-scale angular power spectrum of the diffuse emission seems to obey a power-law with index $2.47 \pm 0.02$ across all energy bands.

    Inspired by the pseudocolor visualization, we decompose the diffuse $\gamma$-ray sky into a ``cloud''-like and ``bubble''-like emission component. The former, tracing the dense, cold ISM, is dominated by hadronic emission processes, while the latter, being morphologically and spectrally distinct, seems to be dominated by leptonic processes in hot, dilute parts of the ISM and outflows thereof. In particular, our findings indicate a preference for IC emission from the \emph{Fermi} bubbles and support scenarios in which the \emph{Fermi} bubbles are explained by hot outflows powered by strong activities in the Galactic center region \citep{YRZ13,C+11,D+11,C-11,C+13}.
    We report further, smaller outflows of a similar population of relativistic particles at other locations.

    The reconstruction of the point-like photon flux qualitatively confirms most of the sources from the second and third \emph{Fermi} LAT source catalog. Quantitatively, we derive the first D$^3$PO \emph{Fermi} catalog of $\gamma$-ray source candidates that comprises $3{,}106$ point sources.
      A more detailed analysis of this catalog is left for future research.

    Finally, we observe $\gamma$-ray emission in the direction of a few galaxy clusters hosting known radio halos. Further analysis is required to shed light on the origin of this emission.

\section*{Acknowledgments}
\footnotesize

    We thank Martin Reinecke, Maksim Greiner, Sebastian Dorn, Dmitry Prokhorov, and Philipp Girichidis, as well as Elena Orlando, and the anonymous referee for the insightful discussions and productive comments. Moreover, we would like to thank Jean Ballet, Johann Cohen-Tanugi, Andrew W. Strong, Roland Crocker, and Stephan Zimmer for their very helpful annotations on the preprint version of this paper. We are grateful to the \emph{Fermi} collaboration for publicly providing their data and data analysis tools.

    This work has been carried out in the framework of the DFG Forschergruppe 1254 ``Magnetisation of Interstellar and Intergalactic Media: The Prospects of Low-Frequency Radio Observations''.
     Furthermore, we thank the ``MaxEnt and Bayesian Association of Australia, Inc.'' for travel support in order to present preliminary results at MaxEnt 2013.

    Some of the results in this publication have been derived using \textsc{NumPy}/\textsc{SciPy} \citep{O06}, \textsc{MatPlotLib} \citep{H07}, \textsc{HEALPix} \citep{G+05}, and especially the D$^3$PO algorithm \citep{SE13} based on the \textsc{NIFTy} package \citep{S+13}.

    This research has made use of NASA's Astrophysics Data System. We acknowledge the use of the Legacy Archive for Microwave Background Data Analysis (LAMBDA), part of the High Energy Astrophysics Science Archive Center (HEASARC). HEASARC/LAMBDA is a service of the Astrophysics Science Division at the NASA Goddard Space Flight Center.

\bibliographystyle{myaa}

\bibliography{fermi.bib}
\normalsize

\begin{appendix}
\section{Data analysis}
\label{app:ana}

\subsection{Data selection}
\label{sec:fermi_data}

    In this work, we analyze the $6.5$ years of observational data taken by the \emph{Fermi} LAT; i.e., data from mission weeks $9$ to $346$ (mission elapsed time $239{,}557{,}417 \, \mathrm{s}$ to $443{,}556{,}512 \, \mathrm{s}$). The data are subject to multiple restrictions and cuts detailed in the following.

    For our analysis we exclusively consider events classified as \texttt{P7REP\char`_CLEAN\char`_V15} in the reprocessed Pass 7 data set. The \texttt{CLEAN} events, which are ``cleaned'' of CR interactions with the instrument, are recommended for studies of the diffuse $\gamma$-ray emission.\footnote{For further details regarding the \emph{Fermi} LAT data products see \url{http://fermi.gsfc.nasa.gov/ssc/data/}.}
    Events with rocking angles (between the LAT boresight and zenith) above $52^\circ$ and zenith angles above $100^\circ$ are excluded in order to suppress contaminations from CRs and the Earth's limb \citep{F09a}.\footnote{The exact filter expression reads \texttt{"DATA\char`_QUAL>0 \&\& LAT\char`_CONFIG==1 \&\& ABS(ROCK\_ANGLE)<52 \&\& ZENITH<100"}.}
    In addition, we apply a (non-standard) cut with respect to the angular distance to the Sun that we require to exceed $20^\circ$. This way, almost all Solar $\gamma$-rays are rejected at the price of reducing the total number of events by less than $3\%$. A similar procedure regarding the moon is conceivable but ignored, because its contribution is negligible.

    The individual events are labeled \texttt{FRONT} or \texttt{BACK} according to whether the photon has been converted in the front or back section of the LAT instrument. We retain this labeling, but combine those data vectors by a direct sum; i.e.,
    \begin{align}
        \bb{d} &= \bb{d}^{\,\texttt{FRONT}} \oplus \bb{d}^{\,\texttt{BACK}} = \left( \bb{d}^{\,\texttt{FRONT}} , \bb{d}^{\,\texttt{BACK}} \right)^\intercal
        .
    \end{align}
    The selected events are binned in nine (logarithmically equally spaced) energy bins ranging from $0.6$ to $307.2 \, \mathrm{GeV}$, cf. Table~\ref{tab:E}. We also apply a spatial binning of the events into all-sky count maps using a \textsc{HEALPix} discretization with $n_\mathrm{side} = 128$, which corresponds to $196{,}608$ pixels with a size of roughly $ 64 \, \mu\mathrm{sr} \approx (0.46^\circ)^2$ each.

    For a proper deconvolution, our analysis has to take the LAT's PSF and exposure into account.
    The instrumental response functions of the \emph{Fermi} LAT \citep{F09a,F09c,F12c}, which are essential therefor, have been improved in the reprocessed Pass 7 release, and are available within the \emph{Fermi} Science Tools. According to our event selection, we make use of the \texttt{P7REP\_CLEAN\_V15::FRONT} and \texttt{BACK} IRFs, respectively.
    These can be assumed to be accurately calibrated, although further improvements, especially at low energies, are under discussion \citep{PF14}.
    Given the IRFs, the exposure $(\varepsilon_{ij}^{\,\texttt{FRONT}} , \varepsilon_{ij}^{\,\texttt{BACK}})$ for each \textsc{HEALPix} pixel $i$ and each energy band $j$ can be retrieved from the data archive. In order to compute $\varepsilon$, the Sun exposure is subtracted from the ``standard'' exposure due to the chosen rejection of potential Solar events \citep{JO13}.
    The PSF of the \emph{Fermi} LAT is a function of position and energy $E$. Its shape varies slightly with spatial translation and sharpens strongly with increasing energy.
    The forward application of the PSF is a linear operation that can be implemented in form of a convolution matrix evaluating the PSF at each pixel center and for each energy band. This matrix is fairly sparse because of the vanishing tails of the PSF, and is computed beforehand.
    The exposure and the PSF define the instrument response operator $\bb{R}$,
    \begin{align}
        R_{ij}(x) &\propto \frac{1}{(E_j^\mathrm{max}-E_j^\mathrm{min})} \left( \begin{array}{c} \varepsilon_{ij}^{\,\texttt{FRONT}} \times \mathrm{PSF}_i^{\,\texttt{FRONT}}(E_j^\mathrm{mid},x) \\ \varepsilon_{ij}^{\,\texttt{BACK}} \times \mathrm{PSF}_i^{\,\texttt{BACK}}(E_j^\mathrm{mid},x) \end{array} \right)
        , \label{eq:response}
    \end{align}
    up to a proportionality constant that can absorb numerical factors and physical units. Notice that this definition does not include a spectral convolution; i.e., no cross-talk between different energy bands is assumed.

    The primary target of our analysis is the physical photon flux $\bb{\phi} = \phi(x)$, which is a function of position $x \in \Omega$. Here, the position space $\Omega$ is the observational sphere, and the position $x$ might be given in spherical coordinates $(\varphi,\theta)$, or in Galactic longitude and latitude $(l,b)$.

    The response operator $\bb{R}$ describes the mapping of a photon flux $\bb{\phi}$ to $\bb{\lambda} = \bb{R} \bb{\phi}$ by a convolution with the IRFs; i.e.,
    \begin{align}
        \lambda_{ij} = \int_\Omega \d x \; R_{ij}(x) \phi(x)
        ,
    \end{align}
    where $\bb{\lambda}$ describes the noiseless (non-integer) number of photons one expects to observe through the IRFs given some photon flux in the sky. This expected number of counts $\bb{\lambda}$ relates to the observed (integer) photon counts $\bb{d}$ by a noise process, which is part of the statistical model detailed in the next section.

\subsection{Inference algorithm}

    The foundation of the analysis presented in this work is the D$^3$PO inference algorithm derived by \citet{SE13} that targets the denoising, deconvolution and decomposition of photon observations.
    Without going into technical details, we briefly review the underlying assumptions and characteristics of the D$^3$PO algorithm in the following.

    The observed photon count data $\bb{d}$ carries information about the astrophysical photon flux $\bb{\phi}$, as well as noise and instrumental imprints. In order to optimally reconstruct $\bb{\phi}$ given $\bb{d}$, we incorporate our knowledge about the actual measurement in a data model that consists of deterministic relations and probabilistic processes. Hence, D$^3$PO is a probabilistic algorithm conducting Bayesian inference.

    We can assume the photon counts to suffer from Poissonian shot noise; i.e., the data entries $d_{ij}$ are the outcomes of statistically independent Poisson processes given an expected number of counts $\lambda_{ij}$ each. Especially in low photon flux regions and at high energies, where the signal-to-noise ratios are worst, using a Poissonian likelihood allows for an accurate noise treatment whereas Gaussian noise approximations often fail.

    D$^3$PO's deconvolution task covers the correction of all effects that trace back to the instrumental response $\bb{R}$. As discussed in Section~\ref{sec:fermi_data}, this response establishes a relation between the astrophysical photon flux $\bb{\phi}$ and the expected counts $\bb{\lambda}$ by taking the IRFs of the \emph{Fermi} LAT fully into account. Since we suppose the IRFs to be thoroughly calibrated (to the best of our knowledge), this relation is deterministic.

    The total photon flux $\bb{\phi}$ consists of many different contributions that can be divided into two morphological classes, diffuse and point-like contributions.
    Diffuse emission, which is produced by the interaction of CRs with the ISM, unresolved point sources and extragalactic background, is characterized by spatially smooth fluctuations. On the contrary, point-like emission is fairly local originating primarily from resolved point sources. The D$^3$PO algorithm reconstructs the total photon flux as the sum of a diffuse and point-like flux contribution; i.e.,
    \begin{align}
        \bb{\phi} &= \bb{\phi}^{(s)} + \bb{\phi}^{(u)} = \phi_0 \left( \e^\bb{s} + \e^\bb{u} \right)
        ,
    \end{align}
    where $\phi_0$ is a constant absorbing numerical factors and flux units, and the exponentiation is applied pixelwise to the diffuse and point-like signal fields, $\bb{s}$ and $\bb{u}$. Those signal fields describe the dimensionless logarithmic flux ensuring the positivity of the physical photon flux in a natural way. Although the algorithm deals with the $\bb{s}$ and $\bb{u}$ fields for numerical reasons, we only regard the fluxes $\bb{\phi}^{(s)}$ and $\bb{\phi}^{(u)}$ in the following as they are physical.

    \begin{figure}[t!]
        \centering
        \begin{tikzpicture}
            [c/.style={circle,minimum size=2em,text centered,thin},
             cr/.style={rectangle,minimum height=2em,rounded corners=1em,text centered,thin},
             r/.style={rectangle,minimum size=2em,text centered,thin},
             v/.style={->,shorten >=1pt,>=stealth,thick}]
            \node(a)at(-2,5)[r,text width=3em,draw,dashed]{$\alpha,\;q$};
            \node(z)at(0,5)[r,draw]{$\sigma$};
            \node(b)at(2,5)[r,text width=3em,draw]{$\beta,\;\eta$};
            \node(t)at(-1,4)[c,draw,dashed]{$\bb{\tau}$};
            \node(s)at(-1,3)[c,draw,dashed]{$\bb{s}$};
            \node(u)at(2,3)[c,draw,dashed]{$\bb{u}$};
            \node(x)at(0.5,3)[cr,text width=5em,draw]{$\;\bb{\phi}^{(s)}+\,\bb{\phi}^{(u)}_{\phantom{u}}$};
            \node(g)at(0.5,2)[c,draw,dashed]{$\bb{\phi}$};
            \node(l)at(0.5,1)[c,draw,dashed]{$\bb{\lambda}$};
            \node(d)at(-1,0)[r,draw]{$\bb{d}^{\,\texttt{FRONT}}$};
            \node(e)at(2,0)[r,draw]{$\bb{d}^{\,\texttt{BACK}}\;$};
            \draw[v](a.south)--(t);
            \draw[v](z.south)--(t);
            \draw[v](t)--(s);
            \draw[v](b)--(u);
            \draw[v](s)--(g);
            \draw[v](u)--(g);
            \draw[thick,double](x)--(g);
            \draw[v](g)--(l);
            \draw[v](l)--(d);
            \draw[v](l)--(e);
        \end{tikzpicture}
        \flushleft
        \caption{Graphical model of the model parameters $\alpha$, $q$, $\sigma$, $\beta$, and $\eta$, the logarithmic angular power spectrum $\bb{\tau}$, the diffuse and point-like signal fields, $\bb{s}$ and $\bb{u}$, the photon fluxes, $\bb{\phi}$, $\bb{\phi}^{(s)}$, and $\bb{\phi}^{(u)}$, and the expected and observed number of photons, $\bb{\lambda}$ and $\bb{d}$.}
        \label{fig:scheme}
    \end{figure}
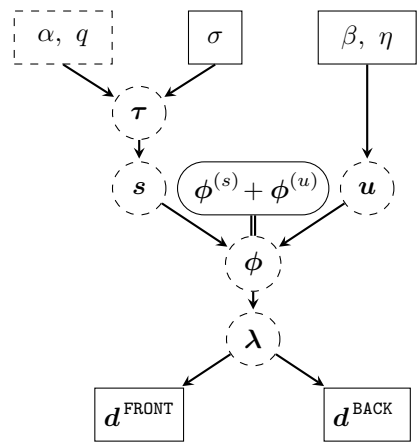

    We can incorporate our naive understanding of ``diffuse'' and ``point-like'' by introducing prior assumptions. Embedding \emph{a~priori} knowledge on the solution, of course, biases the inference. However, priors also remedy the degeneracy of the inference problem as they suppress counterintuitive solutions.

    The diffuse photon flux $\bb{\phi}^{(s)}$, being spatially smooth, is expected to exhibit spatial correlation. Without enforcing concrete spatial features, such as a Galactic profile, we assume $\bb{\phi}^{(s)}$ to obey multivariate log-normal statistics. Assuming, furthermore, statistical homogeneity and isotropy, the underlying covariance is determined by an \emph{a~priori} unknown angular power spectrum. In order to retain a flexible handle on this power spectrum, we further introduce hyperpriors \citep[cf{.} e.g.,][]{EF11,OSBE12,SE13}. We assume a (asymptotically) uniform prior for the logarithmic angular power and a spectral smoothness prior as suggested by \citet{OSBE12}. This introduces scalar model parameters $\alpha$, $q$, and $\sigma$.

    The point-like photon flux $\bb{\phi}^{(u)}$ exhibits strong features that appear to be fairly rare and local.
    We assume $\bb{\phi}^{(u)}$ to follow statistically independent inverse-Gamma distributions described by the model parameters $\beta$ and $\eta$.

    For a detailed derivation and discussion of the probabilistic model D$^3$PO is based on, we refer to \citet{SE13}. An illustrative graphical model of the introduced hierarchy of Bayesian parameters is shown in Fig.~\ref{fig:scheme}.

\subsection{Analysis procedure}

    In theory, we could apply the D$^3$PO algorithm to the whole data set at once. However, it is computationally more efficient to exploit the spectral separability of the response model by applying the algorithm to each energy band individually, cf. Eq.~\ref{eq:response}.
    In order to exploit spectral correlations, we propose to align the priors of the diffuse component after an initial inference run by defining a common angular power spectrum. This yields a three-step analysis procedure detailed in the following.

    \paragraph{Initial inference:}
    The D$^3$PO algorithm is applied to each energy band separately, which can be done in parallel.

    We fix the model parameters with fairly soft constraints, by setting $\alpha = 1$, $q = 10^{-12}$, $\sigma = 10$, $\beta = \tfrac{3}{2}$, and $\eta = 10^{-2}$. The limit of $(\alpha,q) \rightarrow (1,0)$ leads to uniform prior for the logarithmic angular power spectrum $\bb{\tau}$ of the diffuse photon flux \citep{EF11}, but choosing a non-zero $q$ is numerically more stable. The spectral smoothness parameter $\sigma$ is the \emph{a~priori} standard deviation of the second derivative of $\bb{\tau} = \tau_\ell$ with respect to the logarithm of the angular quantum number $\ell$; i.e., $\sigma$ describes the tolerance of deviations from a power-law shape \citep{OSBE12}. The parameter tuple $(\beta,\eta)$ determines the slope and scale of the inverse-Gamma prior of the point-like photon flux. While a slope of $\frac{3}{2}$ is generally applicable, the scale, for which we find $10^{-2}$ fitting, has to be adapted according to the chosen resolution \citep{SE13}.

    D$^3$PO solves an inference problem that is non-linear and, in general, non-convex. To circumvent a dependence on its initialization, D$^3$PO can generate suitable starting values by solving a coarse grained inference problem first. For this purpose only, we provide a binary exposure masking the most prominent point sources.

    \paragraph{Prior alignment:}
       The prior of the diffuse component describes our \emph{a~priori} expectation of how spatially smooth the emission is. If we find diffuse structures of a certain size at one energy band, we can expect to find similar structures at neighboring bands, especially since most diffuse emission processes exhibit power-law-like energy spectra. Thus, we expect significant spectral correlations, in particular for prominent features such as the Galactic bulge, for example. Since the incorporation of a spectral convolution in the response is computationally infeasible, we impose an aligning of the diffuse priors to exploit spectral correlations, nonetheless.

    The diffuse prior is defined by the logarithmic angular power spectrum $\bb{\tau}$.
    As discussed in Sec.~\ref{sec:power}, we find a rough power-law behavior of the power spectra with deviations due to the Galactic disk and finite exposure. Fig.~\ref{fig:power} shows the results of the initial inference, in particular including a power spectrum fit averaged across the energy bands. This average spectrum defines the aligned prior.
    Notice that the apparent excess of small-scale power for high energy bands, comparing inferred and aligned power spectra, remedies potential perception thresholds occurring during the inference \citep{EF11}.

    Afterwards, we also align the diffuse maps within the initially masked regions according to the aligned prior in order to avoid artifacts due to initialization.\footnote{We minimize the prior term, $\bb{s}^\dagger \bb{S}^{-1} \bb{s}$, but only for $s(x | x = (\varphi,\theta) \in \mathrm{mask} \land x \sim (\ell,m) \neq (2\N,0))$ ensuring the preservation of the Galactic profile and the reconstructed field values outside the mask.}

    The alignment of the point-like priors, through $\beta$ and $\eta$, has proven ineffective in tests and is therefore omitted.

    \paragraph{Final inference:}
    The D$^3$PO algorithm is applied to each energy band separately, again. For this run, however, we keep the (aligned) angular power spectrum fixed and provide the aligned diffuse maps as starting values. Hence, the initially used binary mask is not required any more. Notice that a fixed angular power spectrum renders the model parameters $\alpha$, $q$, and $\sigma$ obsolete.

\end{appendix}

\end{document}